\pdfoutput=1 
\documentclass[useAMS,usenatbib]{mn2e}
\usepackage[applemac]{inputenc}
\usepackage[pdftex]{color,graphicx}
\usepackage{amsmath}
\usepackage{amssymb}

\newcommand{\Ha}{H$\alpha$}
\newcommand{\hii}{H{\sc II}}
\newcommand{\hi}{H{\sc I}}

\newcommand{\et}{{\it et al. }}

\def\kms{\mbox{km~s$^{-1}$}}

\def\msun{\mbox{M$_\odot$}}
\def\arcsecpix{\mbox{arcsec~pix$^{-1}$}}
\def\um{\mbox{$\mu$m}}

\def\aj{AJ}                 
\def\apj{ApJ}                 
\def\apjl{ApJL}                 
\def\apjs{ApJS}               
\def\mnras{MNRAS}             
\def\aap{A\&A}                 
\def\aaps{A\&AS}                 
\def\pasp{PASP}		
\def\araa{ARA\&A}
\def\nat{Nature}
\def\pasj{PASJ}

\title[Age patterns in a sample of nearby spiral galaxies]{Age patterns in a sample of spiral galaxies}
\author[M.C. Sánchez-Gil et al.]{M. Carmen Sánchez-Gil$^{1}$\thanks{E-mail:
sanchezg@iaa.es}, 
D.~Heath Jones$^{2}$, Enrique Pérez$^{1}$, Joss Bland-Hawthorn$^{3}$, 
\newauthor Emilio J. Alfaro$^{1}$,  and John O'Byrne$^{3}$ \\
$^{1}$Instituto de Astrofísica de Andalucía, Glorieta de la Astronom\'{\i}a s/n, Aptdo. 3004, 18080 Granada, Spain\\
$^{2}$Australian Astronomical Observatory, PO Box 296, Epping, NSW 2121, Australia\\
$^{3}$Sydney Institute for Astronomy, School of Physics, University of Sydney, NSW 2006, Australia
}

\begin{document}

\date{2011 February}

\pagerange{\pageref{firstpage}--\pageref{lastpage}} \pubyear{2010}

\maketitle

\label{firstpage}

\begin{abstract}

We present the burst ages for young stellar populations in a sample of six nearby ($< 10$\,Mpc) spiral galaxies using a differential pixel-based analysis of the ionized gas emission. We explore this as an alternative approach for connecting  large-scale dynamical mechanisms with star formation processes in disk galaxies, based on burst ages derived from the \Ha\ to far UV (FUV) flux ratio. Images of each galaxy in  \Ha\ were taken with Taurus Tunable Filter (TTF) and matched to FUV imaging from GALEX. The resulting flux ratio provides a robust measure of {\em relative} age across the disk which we discuss in terms of the large-scale dynamical motions. Systematic effects, such as a variable initial mass function (IMF), non-solar metallicities, variable star-formation history (SFHs), and dust attenuation, have been  used to derive estimates of the systematic uncertainty.

The resulting age maps show a wide range of patterns outside of those galaxies with the strongest spiral structure, confirming the idea that star formation is driven one by several processes, largely determined by the individual circumstances of the galaxy. Generally, grand design spirals such as M74, M100, and M51 exhibit age gradients across the main spiral arms, with the youngest star formation regions along the central and inner edges. Likewise, in the dominant star-forming complex of IC 2574 or the ring of M94, the most recent star formation is centrally confined to the regions of star formation activity. In M63 and M74 galaxy-wide trends emerge, suggesting that although most star formation is located along spiral arms, spiral density waves are not the only driver in these cases. We argue that despite appearances, galaxy morphology is not an absolute discriminator of the star formation history of an individual galaxy, nor of the processes triggering it.  We conclude that  \Ha-to-FUV flux ratios are a relatively direct way to probe burst ages across galaxies and infer something of their dynamical histories,  provided that sources of systematics are properly taken into account.
\end{abstract}

\begin{keywords}
galaxies: spiral - galaxies: structure - stars: formation - ISM: HII regions - ISM: general .
\end{keywords}

\section{Introduction}

The origin and persistence of spiral structure in galaxies is still largely an open question, although several plausible alternatives
have been hypothesised. In some cases, spiral structure has variously been attributed to tidally influences of  companions, 
a central rotating  bar, or orbiting dark matter clumps (e.g. Bottema 2003; Dubinski et al. 2008). 
Alternatively, spirals density waves may be self-excited, either through quasi-steady global modes of the underlying disk (Lin \& Shu 1964; Bertin \& Lin 1996),  or as short-lived, recurrent transient patterns from self-gravitational instabilities 
(Toomre 1964, 1990; Sellwood \& Carlberg 1984).  Central to all these hypotheses is the need
distinguish between transient and ongoing star-formation on the very small scales ($<1$\,kpc).

Spatial variation in the current star formation rates (SFR) and star formation history (SFH) across a galaxy can
provide vital clues to its dynamical and secular evolution. Grebel (2000) finds that galaxy mass and environment are important 
factors in determining SFH, and finds a diversity of star formation and enrichment histories among galaxies of the
Local Group, even for galaxies with the same morphological type.
The most common star formation diagnostics use far ultraviolet (FUV), far-infrared (FIR), or nebular recombination 
lines (Kennicutt 1998). UV emission is mainly dominated by OB stars, that  have lifetimes $\lesssim$ 10$^8$ yr 
(Iglesias-Páramo et al. 2004, Kennicutt, 1998). In contrast, the \Ha\ line is strong only in the presence of the most massive, 
hot O stars (masses $>$ 10$M_{\odot}$, ages $<$ 20 Myr) with sufficient UV output to ionize the surrounding hydrogen.
These stars have even shorter lifetimes ($\lesssim$ 10$^7$ yrs; Iglesias-Páramo et al. 2004; Kennicutt 1998)
as they rapidly evolve off the main sequence, making the resulting \Ha\ emission the most  instantaneous probe of star formation.

The ratio of the UV to \Ha\ flux therefore gives a good relative indicator of very recent star formation history (SFH).  As a nascent
star forming region evolves, the \Ha\ line emission declines earlier than the UV continuum, leading to a decrease in the \Ha/FUV ratio. 
With appropriate assumptions about the amount of extinction by dust and the nature of the initial mass function (IMF), this ratio
is a direct indicator of the age of a new star forming region.
Various authors have pointed to other factors that could influence \Ha/FUV flux ratio, such as variations in the 
initial mass function (IMF), star formation history (SFH), and metallicity (e.g. Iglesias-Páramo 2004, Meurer et al. 2009, Lee et al. 2009).
Different levels of dust attentuation between individual HII regions also has a major influence, especially in the highly extincted
FUV region. Clearly, any effort to use \Ha/FUV fluxes needs to account for these systematics insofar as the data allow.

Traditional comparisons between star forming distributions have concentrated on identifying entire \hii\ regions as
individual, distinct sources. Commonly, workers in the field select \hii\ regions by eye
(e.g. Zurita \et 2001; Battinelli \et 2000;  Rozas \et 2000; Knapen 1998; 
Kennicutt 1998). However, such techniques are necessarily subjective and therefore difficult to standardise or reproduce.
The flux thresholds at which regions are defined can vary significantly between different observers, and the ability to
differentiate individual \hii\ regions from more clumpy structures becomes compromised. 
Alternatively, region-finding codes, such as {\it SExtractor} 
(Bertin \& Arnouts, 1996) and REGION (developed by C. Heller) search through 
an image for regions bounded by a chosen flux or luminosity threshold. However, 
such codes rely on the user subjectively placing the boundaries of \hii\ regions, or manually separating \hii\ 
regions that are situated too closely to allow easy discrimination by the software. Furthermore, the threshold flux
is critically dependent achieving good background subtraction which is notoriously difficult in the field of a
nearby galaxy. Clearly by avoiding the need to explicitly define 
\hii\ regions at all, one removes a major source of systematic error.

To this end, we are motivated to use a  pixel-by-pixel differential comparison of star formation regions, analogous to the pixel
Colour-Magnitude Diagrams of Lanyon-Foster et al. (2007), and earlier work by Abraham et al. (1999) and Eskridge et al. (2003).
This technique provides information on the global  SF properties of a galaxy without the need to define \hii\ regions. In principle, 
such measurements are completely characterised by pixel scale, spatial resolution, and flux threshold, and so are therefore fully
reproducible over a diverse sample of galaxies.

In this paper we derive age maps for a sample of six nearby spiral galaxies, using pixel-based methods.
Section  \ref{SData}  describes the data and its reduction, including the methodology of the 
pixel-based mapping technique and corrections for extinction. We also describe our adopted model for the age
calibration and explore its dependency on star formation history, metallicity, and initial mass function. In Section \ref{Results}
we present our age maps and discuss them for each galaxy in turn.. 
In Section \ref{SecRobustness} we analyse the uncertainties and robustness of this method in the context of both random and systematic error. Section \ref{summary} contains a brief discussion and conclusions from this work.

\section{Sample and Data}\label{SData}

\subsection{Galaxy Selection}

The galaxies in our sample were selected across a range of star-forming types based on their orientation and 
proximity. Members were chosen to be as face-on as possible
to mitigate the effects of extinction and scattering, as well as minimising the wavelength shift in \Ha\ due to galactic rotation. 
Galaxies were also chosen with distances nearer than 10 Mpc to allow sufficient  spatial resolution to resolve
individual star forming structures within spiral arms. Finally, all galaxies in the sample have archival
UV images from the Galaxy Evolution Explorer  
(GALEX\footnote{http://galex.stsci.edu/GR2 and GR4}, Martin et al 2005) database and far-infrared 
images from the {\em Spitzer} Infrared Nearby Galaxies Survey 
(SINGS\footnote{http://sings.stsci.edu/}, Kennicutt et al 2003).
The former are used for age dating and to provide estimates of extinction, while the latter are necessary for dust 
extinction correction. The final sample consisted of 6 spiral galaxies, ranging from early to late-type. Two of the spirals
are barred, M100 and IC 2574, and all have spatially-resolved H{\sc I} (Martin 1998). A summary of their main properties is presented in Table~\ref{tab1}.

Optical images were obtained in \Ha\ using the Taurus Tunable Filter (TTF;
Bland-Hawthorn \& Jones 1997) on the William Herschel Telescope (WHT) on 1999 March 4$-$6.
Conditions were photometric with stable seeing of 1.0 arcsec.  TTF was tuned to a bandpass
of width $\Delta \lambda = 20 $\,\AA\  centred at $\lambda_{\rm c} = 6570$\,\AA. The intermediate-width 
R0 blocking filter ($\lambda_{\rm c}/\Delta\lambda = 6680/210$\,\AA) was used to remove
transmissions from all but a single interference order. Table~\ref{tab2} gives details of the observational set-up.
Each galaxy was integrated for 1800 s in \Ha\ and either 120 or 300 s in the continuum (Table~\ref{tab3}).

The UV images come from the Nearby Galaxies Survey of the Galaxy Evolution Explorer mission
(NGS survey, GALEX, Martin \et 2005). This survey contains well-resolved imaging (1.5 \arcsecpix) of 296 and 
433 nearby galaxies for GR2/GR3 and GR4 releases, respectively, in two passbands: a narrower far-ultraviolet 
band (FUV; $\lambda_{\rm eff}/\Delta \lambda=1516 / 268$\,\AA), and a broader near-ultraviolet band 
(NUV; $\lambda_{\rm eff} = 2267 / 732$\,\AA). Archival {\em Spitzer} images for the galaxies of the sample were 
used to provide additional estimates of extinction.

\begin{table*}
\centering
  \caption{Galaxy Parameters$^a$}
  \begin{tabular}{@{}lccccccccc@{}}
  \hline
  Galaxy & RA (J2000) & Dec. (J2000) & Type & Redshift & Dist.$^b$\ & pc/\arcsec$^c$&
Inclin.$^d$\
& Dimensions & M$_B$ \\
 & h~m~s & $^\circ$~$'$~$''$ & &  & (Mpc) & & (deg)&(arcmin) &   \\
\hline
NGC~628 (M74) & 01 36 41.70 & +15 46 59.4 & SA(s)c & 0.002192 & 11.4& 55.27 & 5 &10.5$\times$9.5 & 9.95 \\
IC~2574	&10 28 21.25 &+68 24 43.2 &SAB(s)m &0.000190 &4.0& 19.39 & 77&13.2$\times$5.4 &10.80 \\
NGC~4321 (M100) & 12 22 54.90 &+15 49 21.0 &SAB(s)bc &0.005240 & 16.1& 78.06  & 30&7.4$\times$6.3& 10.05 \\
NGC~4736 (M94) &12 50 53.06 &+41 07 13.7 &(R)SA(r)ab&0.001027 & 4.7& 22.79 & 35 & 11.2$\times$9.1&8.99 \\
NGC~5055 (M63) & 13 15 49.25& +42 01 49.3& SA(rs)bc &0.001681&8.4 & 40.72 & 55.2 &12.6$\times$7.2 & 9.31\\
NGC~5194 (M51) & 13 29 52.71& +47 11 42.6& SA(s)bc pec& 0.00154& 8.1& 39.27 & 20& 11.2$\times$6.9& 8.96\\
\hline
\label{tab1}
\end{tabular}
\begin{flushleft}
$^a$ Sourced from {\it NASA Extragalactic Database} :  -- Position reference -- 20032MASX.C, 1991RC3.9C (M51a); -- Redshift -- Lu \et (1993), Huchtmeier \& Skillman(1998), Rand(1995), Mulder \& van Driel (1993), 1991RC3.9C, Turner \& Ho (1994), respectively for each galaxy.\\
$^{b}$ References : (M74) Tully, 1988, Nearby Galaxies Catalogue; (IC 2574) Karachentsev et al., 2002, A\&A 383, 125; 
(M100) Paturel et al., 2002, A\&A 389, 19;  (M94) Tonry et al., 2001, ApJ 546, 681; (M63) Kennicutt et al., 2003, PASP 115, 928; (M51) Feldmeier, Ciardullo, \& Jacoby, 1997, ApJ 479, 231.\\
$^{c}$ Scale in pc per arcsec in the final \Ha\ images and the age maps plots, where the pixel scale is 1.5\arcsec/px. \\
$^{d}$ Martin \& Kennicutt 2001; Dumke \et 2008; K.T. Chyzy \et 2008.
\end{flushleft}
\end{table*}

\begin{table}
 \centering
  \caption{Instrument set-up for \Ha\ observations}
\begin{tabular}{| l  | c |}
\hline
Date & 1999 March 4 -- 6  \\
Telescope        & WHT \\
Focal station & f/8 \\
Detector & TEK2 \\
Pixel scale (arcsec/pixel) & 0.56 \\
Field of view (arcmin) & 15 \\
seeing (arcsec) & 1.0 \\
tunable filter       & RTTF  \\
\Ha\  filter (R0)    &  $\lambda\lambda $ 668 -- 24\,nm \\
\Ha\  continuum (R1)    &  $\lambda\lambda $ 710 -- 28.5\,nm \\
\Ha\  continuum (R3)    &  $\lambda\lambda $ 760 -- 32.6\,nm \\
\hline
\label{tab2}
\end{tabular}
\end{table}

\begin{table}
 \begin{center}
  \caption{Log of optical and UV exposures (in seconds)}
 \begin{tabular}{@{}lcccc@{}}
\hline
   Galaxy & t$_{H\alpha}^a$ & t$_{cont}^b$ & t$_{NUV}^c$ & t$_{FUV}^d$ \\
\hline
NGC628 (M74) 		& 1800 & 300 & 1644 & 1644 \\
IC 2574				& 1800 & 120 & 1861 & 1861 \\
NGC4321 (M100)		& 1800 & 300 & 2962& 1773\\
NGC4736 (M94)		& 1800 & 300 & 4019& 4019\\
NGC5055 (M63)		& 1800 &120 & 1660 &1660\\
NGC5194 (M51a)		& 1800 & 120 & 2539 & 2539\\
\hline
\label{tab3}
\end{tabular}
\end{center}
\begin{flushleft}
$^a$ \Ha\ image exposure time. \\
$^b$ Continuum image exposure time. \\
$^c$ GALEX near-ultraviolet (NUV) exposure time.\\
$^d$ GALEX far-ultraviolet (FUV) exposure time.\\
\end{flushleft}
\end{table}

\subsection{Data Reduction}\label{DataRed}


The optical tunable filter data were reduced using standard {\sc IRAF}\footnote{IRAF is the  {\it Image Reduction and Analysis Facility}, produced by the National Optical Astronomy Observatories (NOAO).} tasks, as well as some written specifically
for tunable filter data. Bad pixel data were corrected using the {\sc IRAF} {\tt proto.fixpix} task and the median bias level in the overscan regions was subtracted from all frames. Both emission line and continuum images from TTF were flat fielded using a
combination of dome and sky flats. Continuum images used sky flats for the corresponding broad-band filter. 
Night sky rings were present in most TTF images due to the well-known phase variation of transmission wavelength with changing off-axis angle. To remove this effect we assumed fixed ring pattern with radial variation centered on the optical axis and fit and subtracted the background azimuthally, as described in Jones et al. (2002).

Flux calibration of the continuum-subtracted TTF emission-line images was done following Bland-Hawthorn (1995). The TTF bandpass is sufficiently narrow that fluxes can be converted from counts to physical units (erg cm$^{-2}$ s$^{-1}$ \AA$^{-1}$) by multiplying the total counts per second by the CCD gain, and dividing by the reduced telescope area, (which takes the central Cassegrain obstruction into account). This was then divided by the total system efficiency (telescope plus instrument), determined from published spectrophotometric standard star fluxes (Oke 1990). 
Standard stars were observed during each observing run at the same TTF plate spacings as for the science frames.

\begin{figure*}
\includegraphics[width=\textwidth]{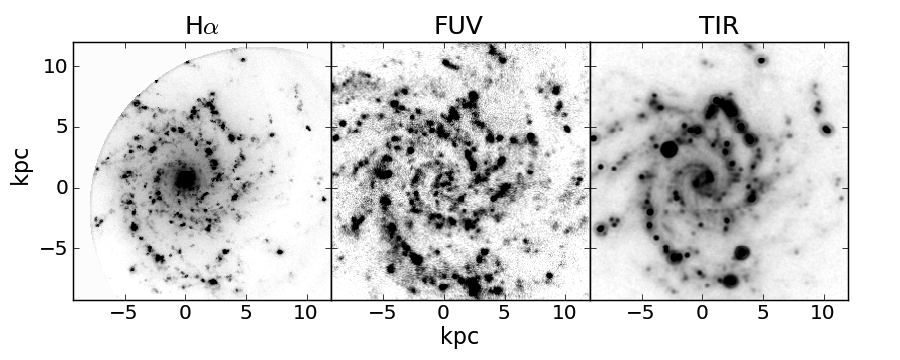}
\caption{Examples of processed frames in  \Ha\ (left), far ultraviolet (FUV; centre), and total infrared (TIR; right) for the galaxy M74. 
The images have been resampled to have identical size, orientation, and pixel scale (1.5''/pix).}
 \label{fig1}
\end{figure*}

We first resampled and aligned our \Ha, UV, and FIR image data onto a common pixel scale and orientation.
Images were trimmed to a common field size, rotated to north-up east-left, and resampled to the lowest scale of all frames  (1.5~\arcsecpix), using the {\sc iraf} tasks {\tt magnify}, {\tt geomap}, and {\tt geotran}. 
This pixel scale is an intermediate value between the better of  \Ha\ (0.56\arcsec) and the worst 
(4.5 and 9\arcsec). 1.5\arcsec is the scale of FUV and 24 mu images. 
It is truth that the actual scale of the resulting TIR image should be 9”, or a bit smaller, but 
not 1.5”. We could be overestimating the ﬂux. But when we resample using 3$\times$3 and 6$\times$6 
binnings, resulting scales of 4.5” and 9” respectively, we check that the age structures 
and gradients remain the same (described in Section \ref{binning}). 
Individual FIR frames were combined into a single total infrared image as described above. The effects of binning the data in this way were investigated and are discussed in Section \ref{binning}.  Galaxy fields were masked to a common area across the \Ha, UV and combined FIR frames.  Those galaxies with prominent bulges (M51 and M74) 
also had these central regions removed, as were foreground Galactic stars and  CCD artifacts.


Galactic extinction was corrected using the Schlegel et al. (1998)\footnote{http://www.astro.princeton.edu/$\sim$schlegel/dust/data/data.html} dust maps for colour excess $E(B-V)$. The extinction, A$_{\lambda}$, at wavelength $\lambda$ was determined from

\begin{equation}
A_{\lambda} \, = \, k_{\lambda} E(B-V)
\end{equation}
where $k_{H\alpha} = 2.54 $, $k_{FUV} = 8.22$ and $k_{NUV} = 8.20$ 
using a Cardelli et al. (1989) extinction curve and R$_V=3.1$. 

Internal extinction for the galaxies in our sample was calculated in a two stages. First, the {\em Spitzer} images
at 24, 70, and 160  \um\  for each galaxy were combined into an image of total far-infrared (TIR) flux, according to
\begin{equation}
F_{TIR} = \zeta_1 \nu F_{\nu}(24 \um)+
 \zeta_2 \nu F_{\nu}(70 \um )+
\zeta_3 \nu F_{\nu}(160 \um)
\label{eqTIR}
\end{equation}
where [$\zeta_1$,$\zeta_2$,$\zeta_3$] = [1.559, 0.7686, 1.347],
(Dale \& Helou 2002). The different plate scales for each frame (1.5~\arcsecpix\ 
for 24 \um, 4.5~\arcsecpix\ for 70~\um, and 9~\arcsecpix\ for 160~\um) 
were all resampled to 1.5~\arcsecpix\ where necessary, identical to that of
the optical and GALEX (FUV and NUV) frames. With this in hand, a value for the 
FUV extinction, A$_{FUV}$, was then derived through Eqn.~2 of Buat et al (2005),
\begin{equation}
A_{FUV} = -0.0333 y^3 + 0.3522 y^2+1.1960 y + 0.4967,
\label{eq9}
\end{equation}
which relates the TIR-to-FUV flux ratio, $y=log(F_{\rm TIR}/F_{\rm FUV})$, also referred to as the 
infrared-excess (IRX). This expression, and the TIR-to-FUV flux ratio, appear to be much a more 
robust and universal tracer of dust extinction than other methods.
As a quantitative dust estimator, it is found to be almost independent of dust and stellar geometry, 
provided that the galaxies are forming stars actively (Buat \& Xu 1996; Buat et al. 1999; Gordon et al. 2000). 
With no available H$\beta$ data, the A(H$\alpha$) extinction was calculated using the relation $A_{FUV} = 1.4 A(H \alpha)$ of Boissier et al. (2005).

\begin{figure*}
\begin{center}
\includegraphics[width=0.45\textwidth]{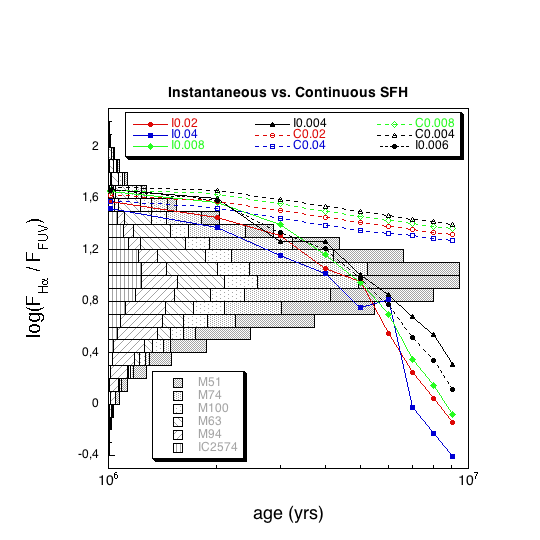}
\includegraphics[width=0.45\textwidth]{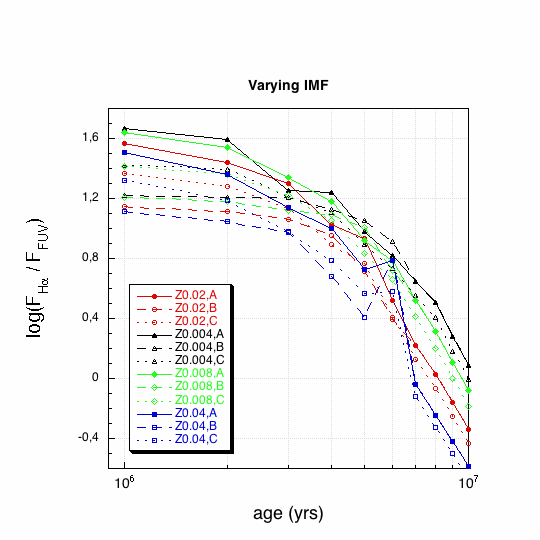}
\includegraphics[width=0.45\textwidth]{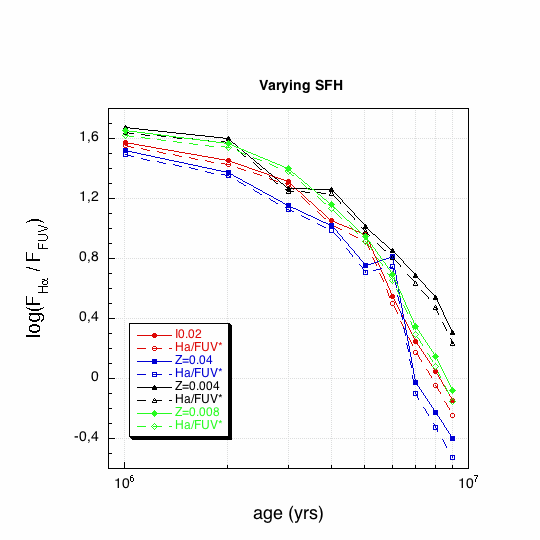}
\includegraphics[width=0.45\textwidth]{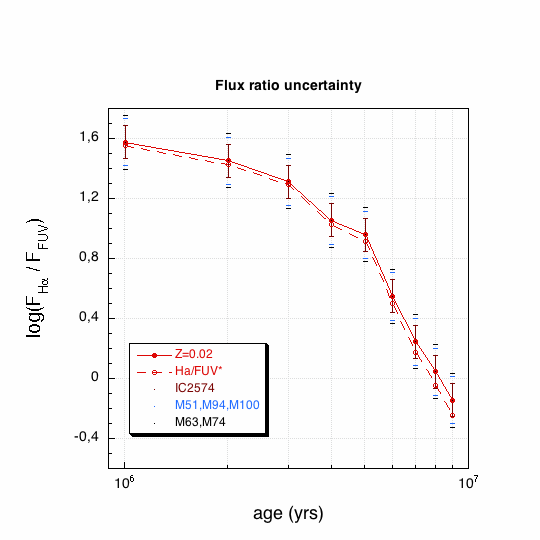}
\caption{The effect of varying SFH, IMF, and metallicity on model \Ha\ to FUV flux ratios, $\log(F_{H\alpha}/{F_{FUV}})$. {\em Upper left:} 
Flux ratio in the case of both instantaneous (dashed lines) and continuous SFHs (solid lines). The models in this panel assume a
Salpeter IMF and range of metallicities: $Z = 0.02$ (red circles), 0.04 (blue squares), 0.008 (green diamonds) and 0.004 (black 
triangles).  The underlying histogram shows the range of measured flux ratios for each galaxy in the sample. {\em Upper right:}  Model 
\Ha\ to FUV flux ratios as a function of varying IMF: (A) Salpeter, $\alpha=2.35$ and $M_{up} = 100M_{\odot}$; (B) truncated 
Salpeter, $\alpha=2.35$ and $M_{up} = 30M_{\odot}$; (C) and Miller-Scalo, $\alpha=3.3$ and $M_{up} = 100M_{\odot}$. In these
cases, the assumed SF law is instantaneous and the metallicities are the same as those in the {\em upper left} panel, and with
the same colour coding. 
{\em Bottom left:} A comparison between an instantaneous SFH with recent 25~Myr-old burst (dashed lines) to a plain instantaneous 
burst (solid lines). The latter are the same solid curves as shown in the {\em upper left} and the same metallicities (and
their colours) have been used.
{\em Bottom right:} A comparison of the systematic model uncertainties with random uncertainty in the flux measurements. 
The instantaneous and (instantaneous$+$recent burst) models are shown (solid and dashed red lines, respectively).
Photometric measurement errors for the galaxies (in three groups) are indicated by the horizontal bars.}
\label{fig2}
\end{center}
\end{figure*}

\subsection{Age calibration}\label{SB99}

Model \Ha\ and FUV luminosities were generated from Starburst99 (Leitherer et al. 1999; V\'azquez \& Leitherer 2005). The 
code can be run with two different star formation modes:  an instantaneous burst or continuous. In choosing our model
inputs we considered three alternatives in the stellar initial mass function, (IMF):
\begin{enumerate}
\item Case A: a Salpeter law with $\alpha=2.35$ and $M_{up} = 100M_{\odot}$ (our reference model),
\item Case B: a truncated Salpeter law with $\alpha=2.35$ and $M_{up} = 30M_{\odot}$, and 
\item Case C: a Miller-Scalo law with $\alpha=3.3$ and $M_{up} = 100M_{\odot}$,
\end{enumerate}
Five metallicities were used for each IMF:   
$Z = $ 0.04, 0.02 (solar, $Z_{\odot}$), 0.008, 0.004 and 0.001. 

The full models cover the age range $10^6$ to $10^9$ yr  in steps of 1~Myr with spectral 
energy distributions  (SEDs) spanning 100 \AA\ to 1 \um\ in wavelength. However, we restricted the 
ages from 1 to 10 Myr as we are only interested in the youngest stellar populations responsible for the \Ha\ and UV output.
We compute our model luminosities as

\begin{eqnarray}
log(L_{H\alpha}) & = & log(EW(H\alpha)) + log(C(H\alpha)) \label{eqLumHa} \\ \nonumber \\
L_{FUV} & = & \frac{\int F(\lambda) S(\lambda) d\lambda}{\int F(\lambda) d\lambda} = 
\frac{\sum_{\lambda=1341}^{1809} F(\lambda) S(\lambda)}{\sum_{\lambda=1341}^{1809} F(\lambda)} \label{eqLumUV} 
\end{eqnarray}

where $EW(H\alpha)$ is the \Ha\ equivalent width (in \AA), and $C(H\alpha)$ is the \Ha\ continuum (measured as the median continuum luminosities at
wavelengths 6550 and 6590 \AA). 
The $F(\lambda)$ term in Eqn.~\ref{eqLumUV} is the GALEX response curve and
$S(\lambda)$ is the luminosity of the SED in units of erg s$^{-1}$ \AA$^{-1}$. 

To assess the suitability and robustness of our reference model we calculated the 
effect of changing various model inputs on the SB99 $F_{H\alpha}/{F_{FUV}}$ ratios (Figure~\ref{fig2}). The upper left
panel shows of Fig. \ref{fig2} shows the ratios for both an instantaneous and continuous SFH, and shows the former
to be much better match to the observed ratios. Like Iglesias-Páramo \et (2004), we find an instantaneous SFH
to be a more sensitive discriminant of the age variations in younger star forming regions. One might naively assume that 
because our galaxies are not starbursts they would be better modeled assuming continuous star formation. 
However, as our pixel approach delineates individual
star forming regions (as opposed to integrating the total star formation across the face of the galaxy), each particular star-forming
region, represented by a pixel in the image, is more accurately regarded as having undergone a stellar burst. Although the 
SFHs of individual regions will be more complex, an  instantaneous starburst is a reasonable approximation 
for localised regions of star formation younger than $10^8$ yr (e.g., Pasquali et al. 2008).

In the upper right panel of Fig. \ref{fig2} we show the effect that changing the IMF has on the \Ha-to-UV ratio for a range of 
metallicities. All three cases of IMF are shown (A, B, and C) which represent a plausible spread of slopes and mass cut-offs.
For each case we apply metallicities of $Z = 0.02$ (solar), 0.04, 0.004, and 0.008. We see that for fixed metallicity, the 
variation between the different IMFs is no more than 0.4~dex for the youngest ages, reducing to $\sim 0.1$~dex beyond
$\sim 6$~Myr. Alternatively, for a given fixed IMF, the variation in \Ha-to-UV with metallicity is at most $\sim 0.5$~dex over
the longest ages, reducing to $\sim 0.2$~dex within the first few Myr.

We would expect that the  \Ha-to-UV ratios are especially sensitive to SFHs that containing a recent burst of star formation 
(say 25~Myr). These would no longer emit in \Ha\ but still carry some output in the FUV given the relatively longer ages 
of the lower mass UV-emitting stars. To investigate this, we compute FUV* which is the FUV flux generated 
from a current burst of star formation combined with a 25~Myr-old burst with identical SFR. 
In the lower left panel of Fig. \ref{fig2} we compare the flux ratio $F_{H\alpha}/{F_{FUV*}}$ (for a SFH with an additional burst)
to the same ratios  $F_{H\alpha}/{F_{FUV}}$ plotted in the upper left (for an instantaneous SFH). The figure shows that
there is negligible difference ($< 0.15$~dex) between either case of SFH, for a fixed given metallicity. We therefore conclude
that recent bursts are much less of an effect on the FUV than one might expect.

The lower right panel of Fig. \ref{fig2} puts the size of this variation due to SFH into the context of the measurement errors.
The red solid and dashed lines show the reference model for both the instantaneous and (instantaneous$+$ recent burst)
cases. Also shown at each point are the $1\sigma$ measurement uncertainties for all of the galaxies. We see that compared
to the observational errors, systematic uncertainty due to the SFH is negligibly small.

We conclude that of all the model inputs, it is choice of IMF that matters most over the first few Myr after an instantaneous burst
of star formation. Metallicity matters more after several Myr have elapsed, and in both cases the range is 0.4 to 0.5~dex at
most, and is comparable with the observation errors. We also conclude that a continuous SFH can be ruled out on the basis 
of our observed $F_{H\alpha}/{F_{FUV}}$ ratios, and that an instantaneous burst gives a more realistic distribution of values. 
While one might expect that the addition of a recent burst of (25~Myr ago) would affect the FUV flux significantly, we find that
this is not the case. Most important of all, we see that while the change in \Ha-to-UV fluxes in {\sl absolute} terms is about 
0.4 to 0.5~dex due to  changing model inputs, the change in {\sl relative} terms (i.e. difference from 1 to 10~Myr) is much
less ($< 0.15$~dex). Therefore, any relative age comparisons are largely immune to the choice of model inputs.

\section{Age Maps}\label{Results}

\begin{figure*}
\includegraphics[width=0.4\textwidth,height=0.4\textwidth]{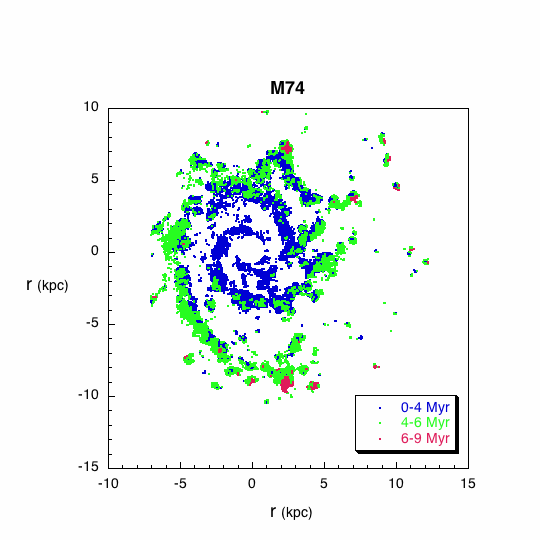}
\includegraphics[width=0.4\textwidth,height=0.4\textwidth]{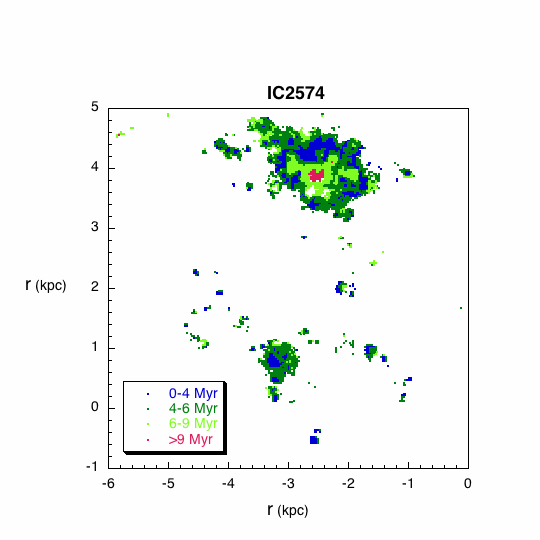}
\includegraphics[width=0.4\textwidth,height=0.4\textwidth]{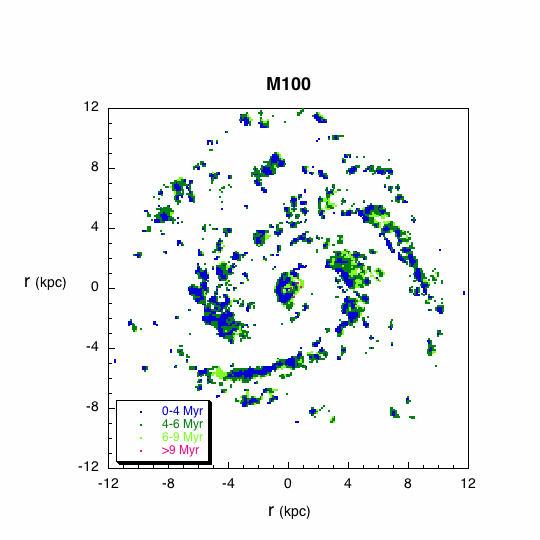}
\includegraphics[width=0.4\textwidth,height=0.4\textwidth]{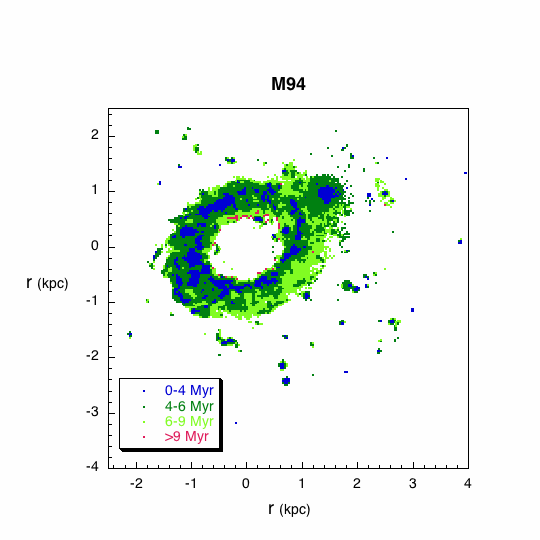}
\includegraphics[width=0.4\textwidth,height=0.4\textwidth]{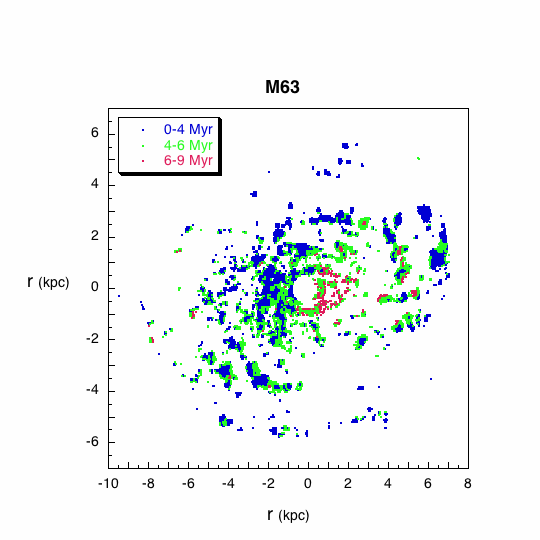}
\includegraphics[width=0.4\textwidth,height=0.4\textwidth]{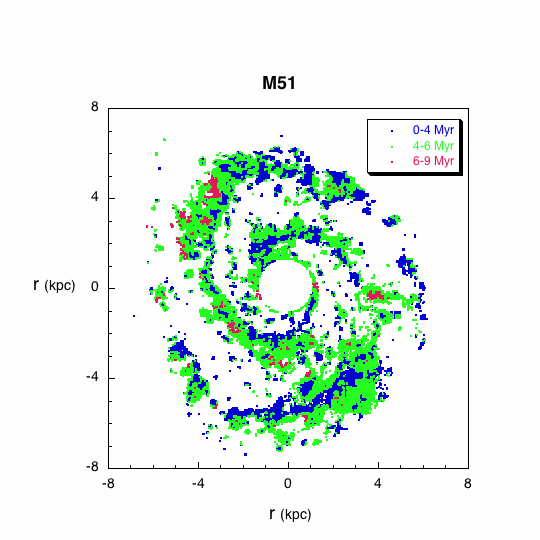}
\caption{Age maps for the galaxies of our sample using the default model, with age denoted by colour as indicated in the key. 
Note that different colour schemes apply to those galaxies with only three age bins (M74, M63, and M51) and the remaining 
galaxies with four. The axes are in kpc and centred on each galaxy. North is up, east to the left.}
\label{fig3}
\end{figure*}

Once the alignment and re-scaling of the images is done, we are in a position to derive calibrated age maps for the sample.
We adopt our reference model which assumes an instantaneous star formation law and Salpeter IMF
($\alpha =2.35$, $M_{up} = 100M_{\odot}$).
In each case we assume solar metallicity, except for IC 2574 where an interpolation between $Z=0.004$ and $Z=0.008$ is applied. 
Radial oxygen abundances from Pilyugin et al. (2004) for all galaxies indicate that  solar metallicity is representative of their mean 
abundances. In the case of IC~2574, Miller \& Hodge (1996) give 12+log(O/H)$\sim$8.15, $\sim30$\% and a solar metallicity 8.69, 
which is why we have adopted the interpolated value $Z=0.006$ for this object.
The flux ratio is calibrated into SB99 model ages at ten intervals of 1 Myr, starting at 1 Myr. This age resolution is dictated by
the internal precision of the flux ratio. 

Figure \ref{fig3} shows the age maps for all galaxies. Each pixel in the map is assigned a range rather than a single valued age, which
is comparable to the photometric flux uncertainties and improves clarity. The maps are contoured in three or four age bins
from the set 0 -- 4 Myr, 4 -- 6 Myr, 6 -- 9 Myr and older than 9 Myr. We discuss each galaxy in turn.

\subsection{M74 (NGC 628)} 

M74 is a well-studied late-type SA(s)c spiral at a distance of
around 11.4 Mpc (Table \ref{tab1}) with a pair of tight spiral arms ($<$ 15 kpc) and no obvious bulge.
 Deep \Ha\ imaging has revealed numerous HII regions in two outer arms at 17 kpc (Leli{\`e}vre \& Roy 2000).
Cornett et al (1994) found the star formation history of this galaxy to vary with radius, consistent with the observation by Natali et al (1992) 
that M74 contains an inner stellar population ($< 1.5$ Holmberg radii $=$ 9\arcmin $=$ 10 kpc) distinct from the outer disc. This separate 
nature of the inner disc is supported by kinematic data (Fathi et al 2007) and Lick indices from the centermost region (Ganda et al 2007).


Figure \ref{fig3} shows our age map for M74. The nucleus of the galaxy is small and bright and was saturated in our \Ha\ maps, and 
so was masked. The age map shows an age gradient from the inner to the outer parts of the galaxy, from very recent to less recent
episodes of star formation, in agreement with previous authors (Cornett et al. 1994, Leli{\`e}vre \& Roy 2000).
Specifically, we find that the \Ha\ luminosity decrease in radius is more pronounced in the inner 5 to 6~kpc, while the UV luminosity
shows a shallower rate of change. Consequently, the \Ha/FUV ratio decreases with radius indicating an age increases in 
the outward direction (Fig. \ref{fig3}).  Cepa \& Beckman (1990) locate the corotation radius at  $\sim 6$ kpc,
although this distance is too close to the outer limit of our data to allow any conclusions about the
effect of corotation to be drawn (see also our discussion of M51 below).  

On more localised scales, the short arm that opens S-SW at 4~kpc shows a clear age gradient across it. 
The outer longer arm, that runs SE-S at $\sim 5$ to 10~kpc, shows a less marked age gradient across its width. 
If the age gradients across spiral arms are a direct product of the spiral density wave, then the dilution of the 
gradient in this southern arm may be related to the weakness of the density wave or the approach to corotation.
As discussed by Efremov (2009), the presence of a shock produced by the spiral density wave, (made visible by dust 
lanes along the spiral arms), is incompatible with the creation of star forming complexes, because of the absence
of visible dust lanes in this arm, despite a chain of complexes along its length. The thickness of the longer arm is basically dominated
by a single age range, and the youngest population located in the inner edge of the arm maps the location of the chain of complexes observed 
in this arm.

\begin{figure}
\includegraphics[width=0.5\textwidth]{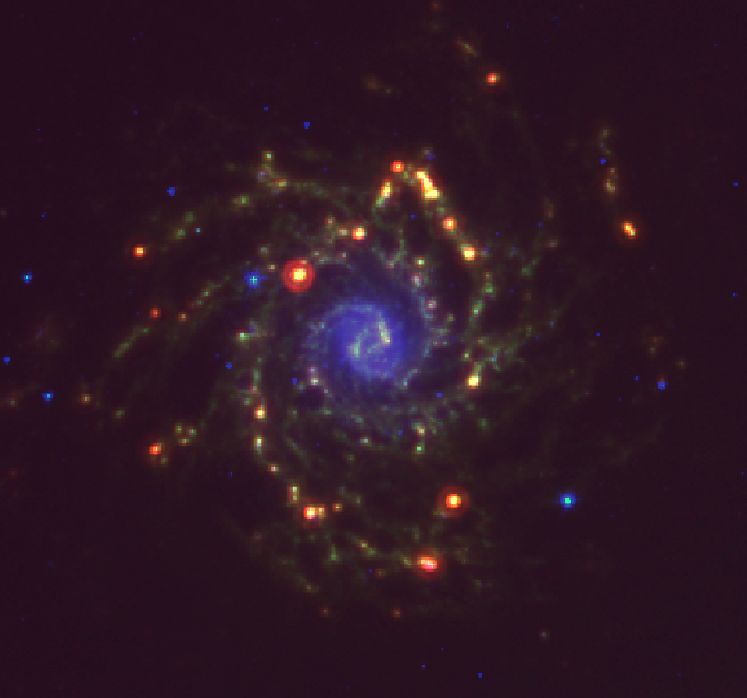}
\caption{False colour composite of M74 in the far infrared, using 3.6 \um\  (blue), 8.0 \um\  (green) and 24 \um\  (red) data.}
\label{fig4}
\end{figure}

An alternative method to gauge the impact of spiral density waves is to check the presence and distribution of an old stellar
population in the far infrared.  Fig.~\ref{fig4}  is an colour composite of M74 in 3.6 \um\ (blue), 8.0 \um\ (green) and 24 \um\ (red) data from the 
SINGS survey\footnote{http://sings.stsci.edu/} (Kennicutt et al 2003). The 24 \um\  image maps the hot dust emission, whereas the 3.6 \um\  image
shows  the old stellar population. This shows how the nucleus, the circumnuclear region, and the inner spiral arms 
of M74 are all dominated by an old population. Cid Fernandes et al. (2004, 2005) and Gonz\'alez Delgado et al. (2004) find 
the majority of the nuclear SED of M74 to be dominated by stars older than 1 Gyr. As is expected for a spiral
density wave, the old stellar populations  are distributed along the inner spiral arms. In contrast, the same 3.6 \um\ light is not seen
along the outer longer southern arm, instead being dominate by the hot dust of star complexes. 
Without the influence of a spiral density wave, as outer arm evolves the stars disperse 
and as a result, the old stellar population is not uniformly distributed along the arm.

A circumnuclear ring of star formation (Wakker \& Adler 1995, James \& Seigar 1999) prompted speculation of a dust-enshrouded bar. Near-infrared 
imaging reveals an oval-shaped central distortion, that could be responsible for the circumnuclear ring, (Seigar 2002) and a weak bar, if any. These 
features are also seen in Fig. \ref{fig4}, and although the nuclear region is masked, the inner regions in the vicinity are within in the youngest age 
range, $<4$ Myr. 

\subsection{IC 2574} 

IC 2574 is a gas-rich dwarf galaxy and a member of the M81 group of galaxies
It is classified as SAB(s)m given its H{\sc I} structure showing two weak spiral arms. In contrast, its \Ha\ and 
UV images show an irregular patchy pattern, which instead suggests a dwarf irregular classification 
(Weisz et al. 2009; Cannon et al. 2005; Walter \& Brinks 1999; Walter et al. 1998).
IC 2574 contains numerous H{\sc I} expanding shells and holes in its interstellar medium. Walter \& Brinks (1999) found 45 large 
expanding shell-like structures in neutral hydrogen, thought to be the combined result of  supernova explosions 
and stellar winds produced by young stellar associations. 

Our age map for IC 2574 (Fig.~\ref{fig3}) is dominated by the giant ($> 1$~kpc)  \Ha\ complex NW of the galactic
centre. This complex is associated with a peak in the H{\sc I} emission (Martimbeau et al. 1994), and with one of the 
prominent expanding supergiant H{\sc I} shells (Walter et al. 1998; Walter \& Brinks 1999; Weisz et al. 2009). 
Weisz et al. (2009) found evidence that stellar feedback created the supergiant H{\sc I} shell with a recent SF episode 
interior to the shell, which peaked $\sim25$ Myr ago, and triggered secondary star formation ($<$10 Myr old) around its rim. 
This corresponds well to the structures seen in our age map of IC 2574 in  Fig. \ref{fig3}. 
As is typical for giant star forming complexes outside of obvious spiral arms, we  find that a younger star forming population 
surrounds an older one located at the centre. Current star formation, as traced by \Ha\ and mid-IR emission, is found along the rims 
of the larger H{\sc I} holes, indicative of  a propagating pattern of star formation as found by other authors (Walter et al. 1998, 
Walter \& Brinks 1999, Cannon et al. 2005,  Martimbeau et al. 1994). S\'anchez Gil et al. (2009) find a similar giant star 
forming complex in NGC 6946 ($\sim700$~pc in diameter), where a central star formation event (a young and massive 
super star cluster) initiated the  expansion of the shell, which swept up the gas and triggered secondary star formation at the rim of the shell.

\subsection{M100 (NGC 4321)} 

M100 (NGC 4321) is a nearly face-on barred spiral in the Virgo Cluster, (16.1~Mpc distant).
The gas kinematics for this galaxy show an intriguing double bar in the gas kinematics, and
García-Burillo et al. (1998) use hydrodynamical cloud simulations to fit a two-bar model simultaneously to
visible, infrared, H{\sc I} and CO data. The two independent systems consist of (i) a nuclear bar 
(with fast pattern speed of $\Omega_f=160$ \kms kpc$^{-1}$ and corotation radius R$^F_{COR}=1.2$ kpc), decoupled
from (ii) an outer bar+spiral with slower pattern speed, ($\Omega_f=23$ \kms kpc$^{-1}$, R$^S_{COR}=8-9$ kpc). 

Our age map for M100 shows an azimuthal gradient for ages less than 10 Myr (Fig. \ref{fig3}). The UV image has 
many bright and extended regions, indicating intense star-formation activity, as a majority of the stars have ages 
younger than 6 Myr. The edge of the age map falls short of the 
outer corotation radius, which  is not seen in our data. The corotation of the fast pattern 
falls within the inner Lindblad resonance (ILR) of the slow pattern,  allowing an efficient transfer of molecular gas 
towards the nuclear region. Both the increased mass inflow rate and the overall higher gas density in the nuclear 
ring should be related to the age gradient detected in the circumnuclear ring of M100 (Ryder et al. 2001, Allard et al. 2006, 
Mazzuca et al. 2008). Allard et al. (2006) find a bipolar azimuthal age gradient from 10 Myr to 50 Myr  using the 
equivalent width of H$\beta$ emission and Ryder et al. (2001) find an azimuthal age distribution 8 to 10 Myr using 
Br$\gamma$ and CO from IR spectroscopy. This consistent picture of the age gradients found through different 
diagnostics (sensitive to different age ranges) indicates that the circumnuclear region of M100 has experienced 
ongoing star formation for at least the past 50 Myr.

The bulk of the circumnuclear star formation events in M100 are best described by
starburst models, with decay time-scales of $\sim$1 Myr (Ryder, Knapen \& Takamiya 2001), as we have used here.

\subsection{M94 (NGC 4736)} 

M94 is an early-type, nearby (4.7 Mpc), spiral with a bright inner region composed of a bright circular bulge 
surrounded by a ring of active star-forming regions,  about 1 kpc from the center (Wong \& Blitz 2000). 
This ring is the dominant feature in the age map of M94 (Fig. \ref{fig3}) and most star formation in the galaxy is 
circumscribed by this region of young hot stars. 
Figure~\ref{fig3} shows that the most recent star formation is innermost around the N sector of the star forming ring but
moves closer to the inside edge around its S sector. The age map also shows an additional star forming complex
on the outer NW edge of the ring.  In the UV image a dim bulge and short 
spiral arms are visible, with the bright ring around the center of the galaxy. In the \Ha\ image, the spiral arms were
too faint to be seen and the bulge was saturated and so was masked out.

A clear age gradient is observed across the ring, with the younger stars in the middle, surrounded by older ones both in 
the inner and outer rims. This young burst of star formation in the ring is also seen in Fig.~3 of Trujillo et al. (2009): 
the color luminosity profiles presented by these authors clearly show that the continuum is enhanced locally at the 
distance of the ring ($\sim47$\arcsec) in the ultraviolet and blue bands, but not in the redder bands. Furthermore, the 
PAH emission bands (5.8, 8.0 and 24$\mu$) are locally enhanced in the mid-infrared (MIR) luminosity profiles, but not the
NIR or redder FIR bands corresponding to the older stars or colder dust. A cursory glance at the MIR Spitzer spectra 
indeed show  the presence of strong PAH emission (see also Fig. 6 of Trujillo et al. 2009). 


The star formation rate computed in the inner part of M94
 is $\sim0.1$ \msun\ yr$^{-1}$kpc$^{-2}$, and can be deduced from the data in Boissier et al. (2007, their Fig.~9.39, 
 for {\hi\}+H$_2$) and Trujillo et al (2009, for their 'bulge' inner region, from FUV data). From our {\Ha\} image the 
 distribution of pixel-wise SFR is 0.1 \msun\ yr$^{-1}$kpc$^{-2}$ for pixels in the 85 percentile band.
Wong \& Blitz (2000) analyse CO, H{\sc I}, and \Ha\ data for this ringed galaxy, and conclude that the star formation rate
is not determined solely by available gas mass, but rather that large-scale dynamics play a significant role in organizing 
and possibly triggering star formation.

M94 is not completely axisymmetric but has a large-scale, broad oval that may affect its internal dynamics (Kormendy \& Kennicutt 2004). 
H{\sc I} synthesis observations (Bosma, van der Hulst, \& Sullivan 1977) and optical and near-infrared imaging 
(M\"{o}llenhoff, Matthias, \& Gerhard 1995) have demonstrated that the disk of the galaxy is nonaxisymmetric, and hence
 the rings may occur at the inner and outer Lindbland resonances (ILR and OLR) of an oval potential (Gerin, Casoli, \& Combes 1991).
Mulder \& Combes (1996) found that the potential of M94 can be modeled well with an oval component at 
intermediate radii. These hydrodynamic simulations suggested that the rotating oval is indeed responsible for the
 'rings' formation, revealing the formation of two rings, at 1.5 and 10 kpc radius, which correspond to the ILR and the 
 OLR, with a pattern speed of 40 km s$^{-1}$ kpc$^{-1}$. In addition, M\"{o}llenhoff, Matthias, \& Gerhard (1995) have 
 suggested that the inner ring may coincide with the OLR of a central stellar bar, $\sim 30${\tt ''} in extent, seen in optical 
 and near-infrared isophotes. The rotation curve is consistent with a scenario in which the ring corresponds to the OLR 
 of the nuclear bar and the ILR of the large-scale oval distortion. Such a coupling of nested bars may contribute to the 
 accumulation of gas in the ring (Wong \& Blitz 2000). Chy$\dot{\rm{z}}$y \& Buta (2008) have demonstrated that M94 
 has highly symmetric and strong magnetic spiral fields, not clearly associated with the shape of the distribution of 
 star-forming regions or with spiral density waves.

\subsection{M63 (NGC 5055)} 

M63 is a nearby, spiral galaxy 
and one of the prototype arm-class 3 flocculent galaxies (Elmegreen \& Elmegreen 1987).
It shows a regular, two-arm spiral structure to a radius of 4 kpc in the near-infrared (Luo et al. 2007) and
has been studied extensively across the electromagnetic spectrum. Luo et al. (2007) studied the
{\it Chandra} properties of low-and-high mass x-ray binary populations, and
concluded that the disk has undergone recent, strong starbursts that significantly increased the
population of high-mass x-ray binaries (HMXB). 
The galaxy has a flocculent appearance in \Ha\ and is significantly brighter on the E side than the W.  
Ongoing starbursts are also observed in the  LINER nucleus of M63, which is UV-bright and surrounded 
by luminous young star clusters, showing clear stellar absorption signatures (Maoz et al. 1998; Leitherer et al. 2002). 
In our \Ha\ data the nucleus is saturated and was masked. 

The age map for M63 in Fig.~\ref{fig3} shows an E-W age gradient running across the face of the galaxy.
This gives it a peculiar bipolar age distribution, in which the pixels in the older age bin (6 -- 9 Myr) are mostly 
located in the W half of the galaxy, the pixels in the intermediate age bin (4-6 Myr) are more evenly distributed 
throughout, and the pixels corresponding to the youngest age bin are concentrated towards the east. Within the
inner $\pm3$ kpc, the W side is dominated by an older population (6 -- 9 Myr) while the E side is dominated 
by one younger than 4 Myr.

M63 has been studied in $^{12}$CO (J$=$1-0) by Tosaki et al. (2003) and in H{\sc I} by Battaglia et al. (2006). 
These authors note the regularity and symmetry of the galaxy, but a mild lopsidedness is noticeable, both in
the distribution and the kinematics of the gas. These results point at two different dynamical regimes: an inner
region dominated by the stellar disk and an outer one, dominated by a dark matter halo offset with respect to
the disk. 
They suggest that cloud formation and 
the ensuing star formation occur both in the arm and the interarm regions due to enhancement of gas by local
fluctuations, since no clear systematic offset between the molecular gas and \hii\ regions is found, as neither are 
Giant Molecular Associations. 

The inner $<3$ kpc in M63 is conspicuously similar in its spatial structure to the dust distribution and to the velocity
field. We have investigated several possible astrophysical causes for an asymmetric age distribution at larger radii.
We do not find any possible mechanism that would relate the age map to the velocity field. We have also looked for a
possible relation with the extinction correction, based on the appearance of the dust distribution in Fig. 12 of La Vigne
et al. (2006). We correct for extinction by means of the FUV/TIR ratio. When this is plotted against the three age bins
that span the age range under consideration, the difference of median value of this ratio is only 0.1 between the first
and third age bins. This small difference translates to a negligible age increment, and thus the extinction pattern can
be rejected as the source for the peculiar age pattern. In the third place we have looked into the polarization pattern
as shown in Fig. 3 of Knapik et al. (2000). The polarization vectors are very uniformly distributed in the region where
we find the age pattern, so this does not seem to be related either. We have also looked into the relationship with the
molecular distribution, as presented by Thornley \& Mundy (1997), but the CO distribution is fairly symmetrical both
sides of the bulge at $\pm2$ kpc, so it does not provides any further information. Thornley \& Mundy (1997) present
a $K$ image in their Fig. 3; and in this image there is a hint of the W side in the inner 2 kpc being brighter than the E
side, what is consistent with the age difference we find in the age map of M63.

\subsection{M51 (NGC 5194)} 

M51 is grand-design spiral galaxy interacting with the nearby NGC 5195 and a member of the same group as M63. 
It is a metal-rich galaxy, $12+$log(O/H)$\sim 8.7-8.9$ (Bresolin \et 2004), with a weak metallicity gradient 
as a function of galactocentric distance (Zaritsky, Kennicutt, \& Huchra 1994). Because of its large apparent
size and luminosity it is possible to observe a wealth of detail in its arms and spiral structure. 

The age map for M51 in Fig.~\ref{fig3} is dominated by a young population of stars ($<6$ Myr). It shows an 
age structure with gradients across the arms, with the younger stars towards the inner edge while the stars 
in the older age bin are located towards the outer edges. This age pattern is expected if star formation is 
triggered by gas shocked by the pass of a spiral density wave (Roberts 1969; Mart{\'i}nez-Garc{\'i}a \et 2009). 
If this is the case, we should observe the inverse age pattern outside co-rotation --- that is, the youngest 
population preferentially located radially in the outer  side of the arms. Determinations of the co-rotation radius 
for M51 are in the range 126\arcsec--161\arcsec\  (Vogel et al. 1993; Elmegreen et al. 1992; Knapen et al.
1992; Garc\'ia-Burillo et al. 1993). These estimates  correspond to a distance of $\sim4.9-6.3$ kpc; unfortunately 
our data falls just short of this range, so we are not able with this data set to confirm whether the age pattern
switches at co-rotation.

Calzetti et al. (2005) present a panchromatic view, UV to FIR, of the star formation in M51 and the impact of 
dust extinction. In their Fig. 1 (left) they present an $(R,G,B)=$ (24\um, \Ha, FUV) image that further 
supports the age pattern across the arms that we find here. What the image shows is the same age pattern 
with the dust+H$\alpha$+FUV located towards the inside of the arms, while the outside is dominated by the 
FUV alone. As we have argued above, Calzetti et al. also state that the UV emission traces predominantly 
the evolved, non-ionizing stellar population, up to ages $\sim50-100$ Myr. Thus the age pattern that we already 
see for the ongoing star formation, $<10$ Myr, can be extended up to the recent past non-ionizing star formation, 
that follow in the outer part of the arms. What we are witnessing here is a clear age pattern along the spurs 
in the arms of the galaxy.

Scheepmaker et al. (2009) have studied the age distribution of 1580 resolved star clusters in M51 from HST 
UBVI photometry, and their spatial relation to the \Ha\ and 20 cm radio-continuum emission. The positions 
of the youngest ($<10$ Myr) clusters show the strongest correlation with the spiral arms, \Ha, and the 
20 cm emission, and these correlations decrease with age (their Figs. 16 and 17). The azimuthal distribution 
of clusters in terms of the kinematic age away from the spiral arms indicates that the majority of the clusters 
formed 5--20 Myr before their parental gas cloud reached the centre of the spiral arm. The authors divide the 
sample in three age ranges, log(age) $<7.0, 7.0-7.5, >7.5$, and find (their Fig. 13) that the oldest clusters have 
older kinematic ages (10--15 Myr) compared to the intermediate age and to the youngest clusters.

\section{Reliability and Robustness}\label{SecRobustness}

The photometric uncertainty on the F$_{H\alpha}/$F$_{FUV}$ flux ratio is dominated by the error
on the FUV flux (15 to 25\%) and not H$\alpha$ ($<5\%$). An additional error term comes
from the use of Equation \ref{eq9} for the FUV extinction correction, which depends on the
uncertainties in the FIR. The uncertainties 
for the 24$\mu$m, 70$\mu$m and 160$\mu$m fluxes are 4\%, 7\% and 12\% respectively (Dale et al. 2007),
implying 6\% uncertainty overall in $F_{\rm TIR}$ and 13\% in the $F_{\rm TIR}/F_{\rm FUV}$ ratio.
With the application of the extinction correction the overall uncertainty in the
F$_{H\alpha}/$F$_{FUV}$ flux ratio is more like 28\%.


Photometric uncertainties aside, the age-dating technique is subject to a number of 
potential sources of systematic error. In this section we test the robustness of our results to systematics
by varying the assumptions we have made over a plausible range of values. Specifically, the effects we examine are
(i) the lowest limit on cluster mass allowable for our assumptions on ionizing flux, (ii) the effect
of changing the spatial bin size, and (iii) the effect of changes to the metallicity and IMF assumptions
in the model. We deal with each of these in turn.

\subsection{Lowest Threshold Pixel Mass for a Fully Sampled IMF}\label{ApplSB99}


\begin{figure*}
\includegraphics[width=0.48\textwidth]{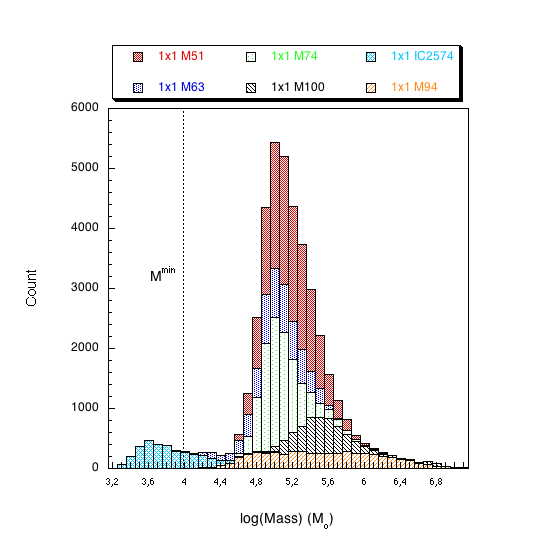}
\includegraphics[width=0.48\textwidth]{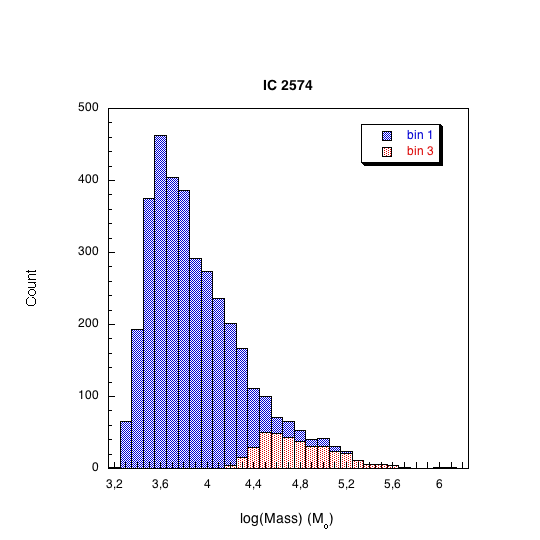}
\caption{{\em Upper left:} The distribution of pixel stellar masses for all the galaxies of the sample, 
assuming the default $1\times1$ binning. The dotted line marks the limiting mass, $M^{min}$, corresponding 
to the Lowest luminosity Limit of Cerviño et al. (2003). {\em Upper right:} The same for IC 2574 only in two instances 
of binning: $1 \times 1$ (blue shading) and $3 \times 3$ (red).}
\label{fig5}
\end{figure*}


When stellar clusters are modeled by simple stellar population synthesis (SPS) techniques  
over a relatively small spatial extent, incomplete sampling of the initial mass function is a concern.
Cerviño et al. (2003) have established a threshold called the Lowest Luminosity Limit (LLL),
above which, synthesis models can be applied to stellar clusters to relate ionizing flux to newborn stellar mass.
The mass of a stellar system with a completely sampled IMF and luminosity equal to the LLL 
is $M^{min}$. The minimum mass depends on the metallicity, the age,  the evolutionary tracks and model 
atmospheres used for the particular SPS model and Cerviño et al. (2003) publish these boundaries for SB99,
as well as for their own models.

We have calculated the stellar masses of our pixels and compared with the values of $M^{min}$ given by Cerviño et al. (2003). 
Mass is calculated based on the FUV luminosity, as a more reliable estimator of the underlying stellar 
population. The pixel stellar mass is a lower limit derived by comparing the linearly scaled extinction-corrected 
observed L$_{FUV}$ with the highest expected value from SB99, for a stellar population mass of $10^6$ M$_{\odot}$ 
at the youngest cluster ages. 
The $M^{min}$ for the SB99 models (with metallicities $Z=0.02$, 0.004 and 0.008) is less than $10^4$ \msun\ for the 
young stellar population, $<9$ Myr, as it is in our case (cf. Cerviño et al. 2003, their Fig. 5).

Figure \ref{fig5} shows the pixel mass distributions for each galaxy compared to the $10^4$ M$_{\odot}$ mass limit,
using the $1\times 1$ binning applied to all galaxies for our main results. Individual pixels range from $\sim 10$ to 50 pc
on a side (Table \ref{tab1}). Save for the irregular IC 2574, all galaxies have pixel stellar masses greater than 
$10^4$ M$_{\odot}$, and therefore lie beyond the threshold for incomplete sampling of the IMF. 
IC 2574 is less massive than the rest of spiral galaxies of the sample, as it is usual for irregular galaxies. 
In the right panel of Fig. \ref{fig5} we show the effect of binning the IC2574 data by $3 \times 3$ which puts the
spatial sampling of IC 2574 above the threshold (cf. Fig. 5 of Cerviño et al. 2003). In the following section we show
that the age map for IC 2574 is the same irrespective of whether $1\times1$ or $3 \times 3$ binning is used.

\subsection{Spatial Binning Scale} \label{binning}


\begin{figure*}
\includegraphics[width=0.45\textwidth]{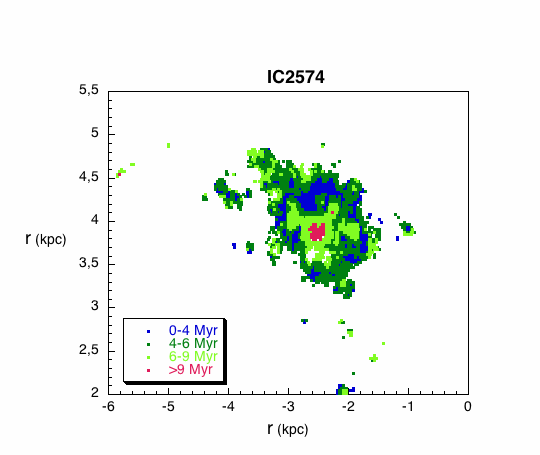}
\includegraphics[width=0.45\textwidth]{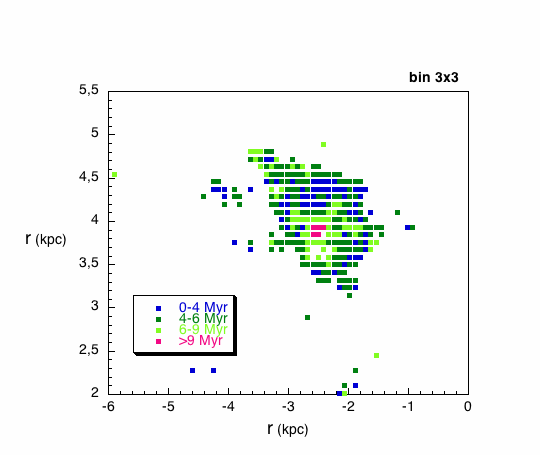}
\includegraphics[width=0.45\textwidth]{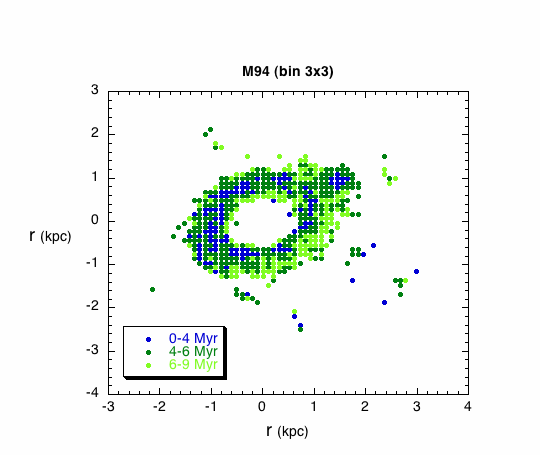}
\includegraphics[width=0.45\textwidth]{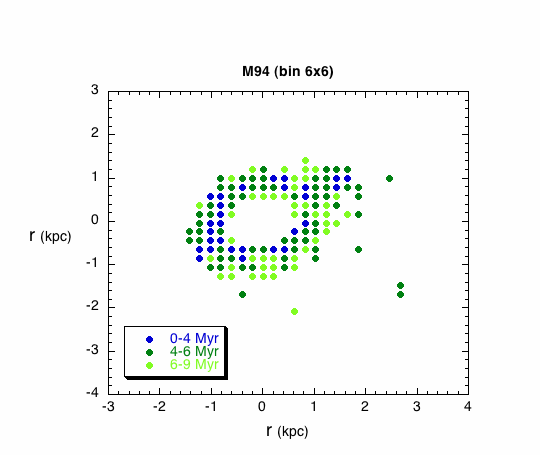}
\includegraphics[width=0.45\textwidth]{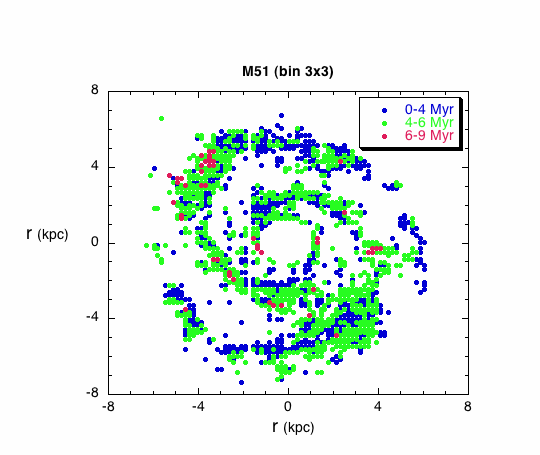}
\includegraphics[width=0.45\textwidth]{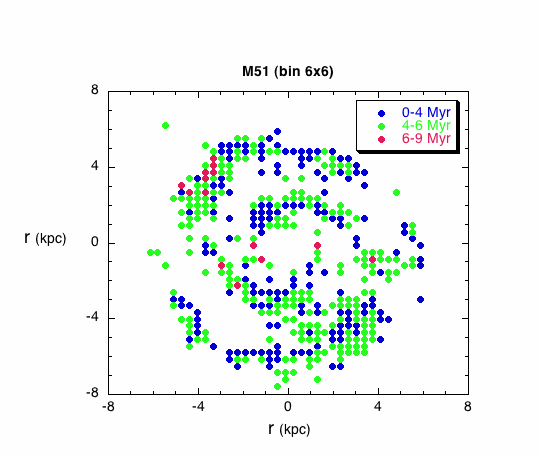}
\caption{Age maps for IC2574, M94, and M51 derived by binning the data to coarser spatial scales. {\em upper panels:}
A zoomed view of the northeastern stellar complex of IC 2574 at its original scale and with $3 \times 3$ binning.
{\em middle panels:} M94 showing cases of $3 \times 3$ and $6 \times 6$ binning. {\em lower panels:} Same binning scales
for M51.The pixel scales are 4.5''/pix and 9''/pix for $3 \times 3$ and $6 \times 6$ binning respectively.}
\label{fig6}
\end{figure*}


We now examine the issue of whether the \Ha/FUV age calibration is valid on spatial scales
other than those used for our results  ($1 \times 1$ binning). This is to verify whether the amount of
\Ha\  and FUV flux in an individual pixel reflects the number of ionizing O and B stars within
the same pixel. Our results should be independent of the geometry of individual HII regions
on the scales at which we sample the galaxies.

Figure \ref{fig6} shows coarser resampling of three galaxies (IC 2574, M94, and M51) using $3 \times 3$ and
$6 \times 6$ binning, (to compare to the $1 \times 1 $ binning used in Fig.~3). The three galaxies we have chosen
are representative of the range of morphologies encountered in the full sample: spiral arms, a stellar ring, and a
giant star forming complex. In the case of IC 2574, we concentrate
on the giant northeastern star forming complex at the original sampling and  $3\times3$ binning; the size of this star forming
region is too small to justify the $6 \times 6$. 

As Fig. \ref{fig6} shows, the age structures and gradients remain the same irrespective of binning scale used. This 
demonstrates the robustness of the pixel-based technique against systematics due to sampling. Furthermore,
the resampling of the 24$\mu$m (1.5 arcsec), 70$\mu$m (4.5 arcsec) and 160$\mu$m (9 arcsec) {\em Spitzer} images to a 
common 1.5 arcsec plate scale has not affected our estimate of the total infrared flux nor its use in the dust corrections.

One of the galaxies in our test set, M51, has several independent flux measurements of its HII regions in the literature,
against which we can test the robustness of our own measurements to the effects of binning.
Scoville et al. (2001) show the distribution of HII diameters in M51 in their Fig. 13. Our smaller 1.5 arcsec pixel size
corresponds to 70 pc (using the distance in Scoville et al), a size that encloses some 95\% of all M51 HII regions.
So even at the limit of our smallest sampling scale we have pixels that wholly contain the vast majority of HII regions
in that galaxy.

\subsection{Metallicity and IMF Variation}\label{UNCERTAINTYMAPS}

The ages we have derived depend on assumed values for metallicity, as well as the adopted
initial mass function (IMF). We have calculated age maps from the full range of plausible metallicity
and IMFs and used the spread in resulting ages to assign a measure of confidence to the age maps
we have derived earlier. We also incorporate flux measurement uncertainties into the estimate.
Each pixel age is computed for a grid of possible models spanning metallicity and IMF, and
the mode of all ages, is chosen as the reference value. 
We define the confidence of our age estimation as the probability that the age
stays with the same range (as set by the modal age) among all possible models.
The confidence interval is calculated from the $3\sigma$ 
spread in computed ages from the modal age. Age values are binned in intervals of 0--4, 4--6, 6--9,
and $>9$ Myr, for the sake of clarity. 

Fig. \ref{fig7} shows the confidence and uncertainty maps for M74, as well as the lower pixel threshold mass 
(as calculated in Section \ref{ApplSB99}). We also show the modal age map across all computed ages, which
 is a useful point of reference for
our final mass maps presented in Sect. \ref{Results}.  Figures \ref{fig7} -- \ref{fig12} show the same
statistical estimates for the other galaxies in our sample.
The average confidence across galaxies is 64-70\%, with around 80\% of pixels having confidence values in 
excess of 50\%. The average uncertainty in age is $1.4\pm0.5$ Myr, well below the youngest ages (4 Myr) across
all possible models. The youngest regions have the highest confidence 
values (in excess of $>80\%$) and the lowest uncertainties. Examples of this are the inner ring in the stellar
complex of IC 2574, the circumnuclear region of M94, or in the spiral arms of M100.
Furthemore, comparing the mass maps with the conﬁdence maps we do not ﬁnd any
effect of the mass in the age assignment. 
In M100  for example, there is a clear correspondence between the regions of high confidence with the most
massive ones. But in M74 and M94 we find that there are regions of intermediate-high masses with lower confidence, 
probably due to a larger variability from the models.

\begin{figure*}
\centering
\includegraphics[width=0.45\textwidth]{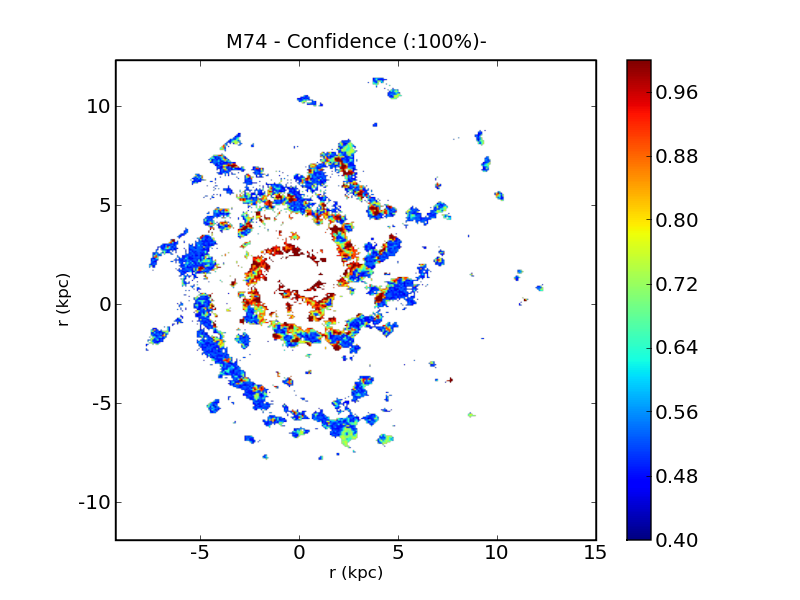}
\includegraphics[width=0.45\textwidth]{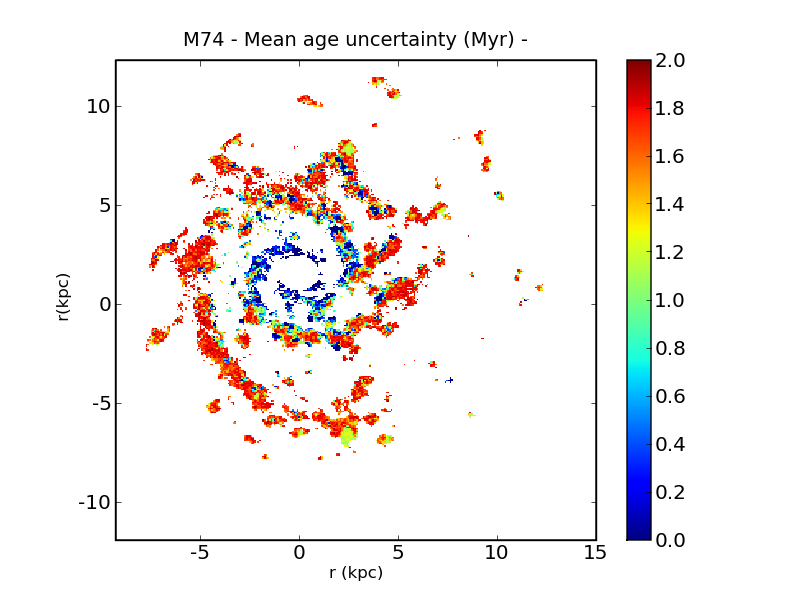}
\includegraphics[width=0.45\textwidth]{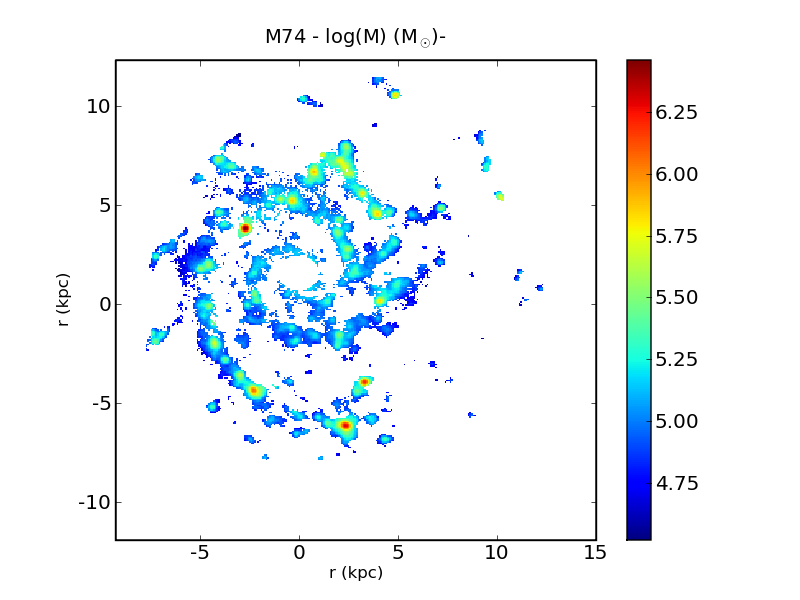}
\includegraphics[width=0.45\textwidth]{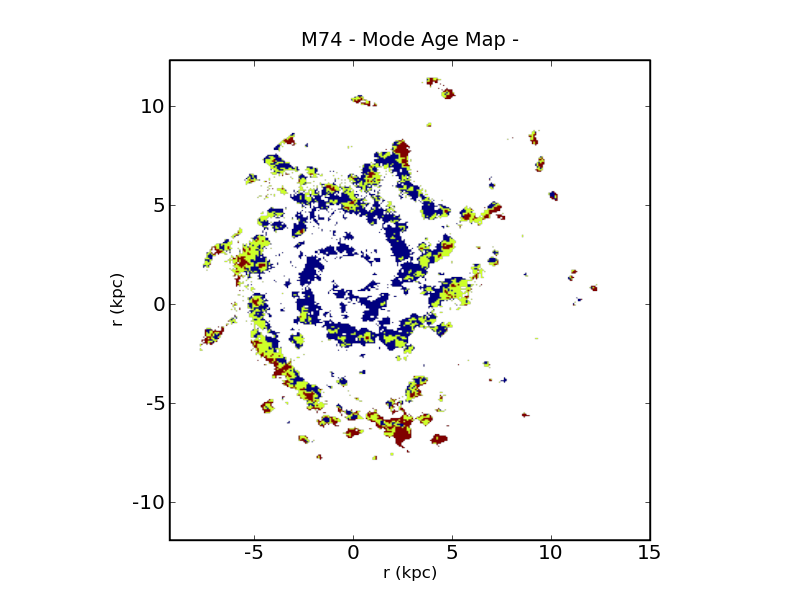}
\caption{Statistical tests of the age map derived for the galaxy M74.
{\em Upper left:} Confidence map of age assignments, expressed as a probability and calculated from the full 
range of plausible SB99 models. It also includes flux ratio uncertainties. 
{\em Upper right:} Map of age uncertainty, expressed in Myr. 
{\em Lower left:} Map of lower pixel threshold mass, calculated in Sect. \ref{ApplSB99}.
{\em Lower right:} Modal age map, in which each pixel age is the mode of all ages derived from the full range of 
plausible SB99 models and flux ratio uncertainties. The pixel scale is 1.5''/pix}
 \label{fig7}
\end{figure*}

\begin{figure*}
\centering
\includegraphics[width=0.45\textwidth]{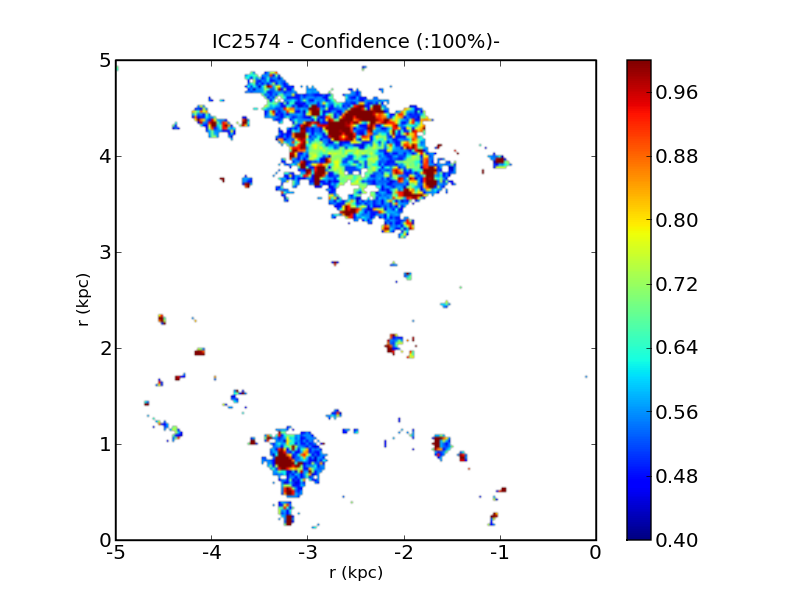}
\includegraphics[width=0.45\textwidth]{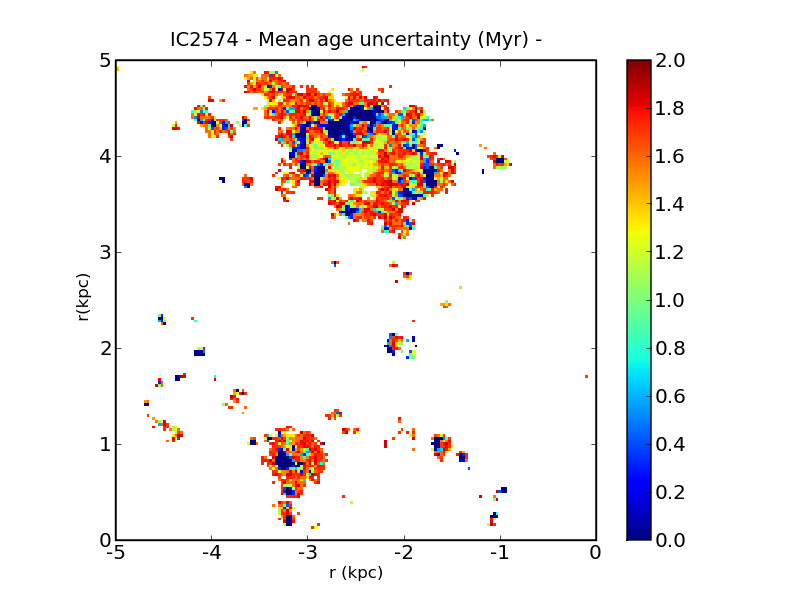}
\includegraphics[width=0.45\textwidth]{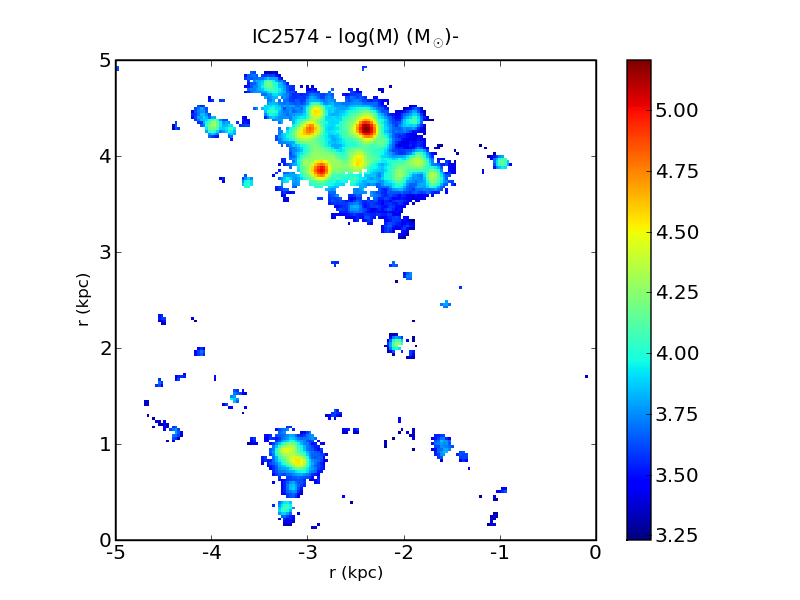}
\includegraphics[width=0.45\textwidth]{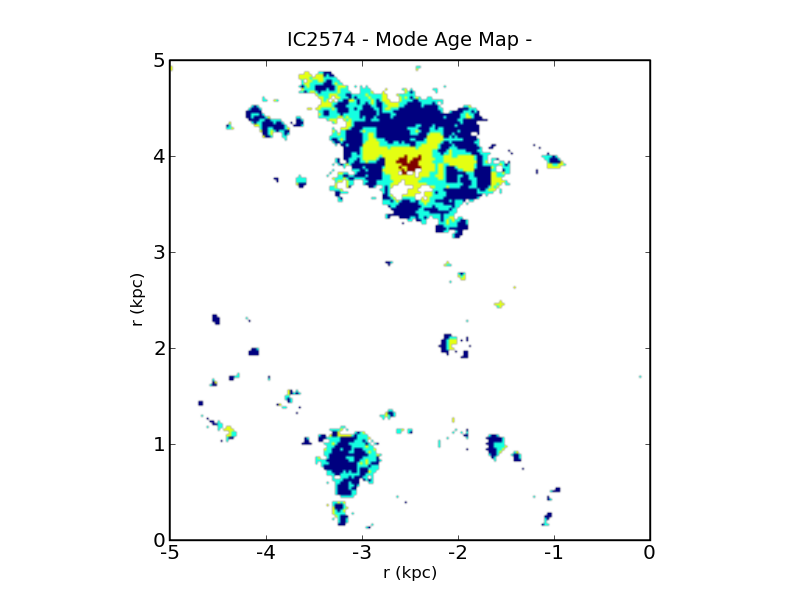}
\caption{Same as Fig. \ref{fig7} for the galaxy IC 2574.}
 \label{fig7}
\end{figure*}

\begin{figure*}
\centering
\includegraphics[width=0.45\textwidth]{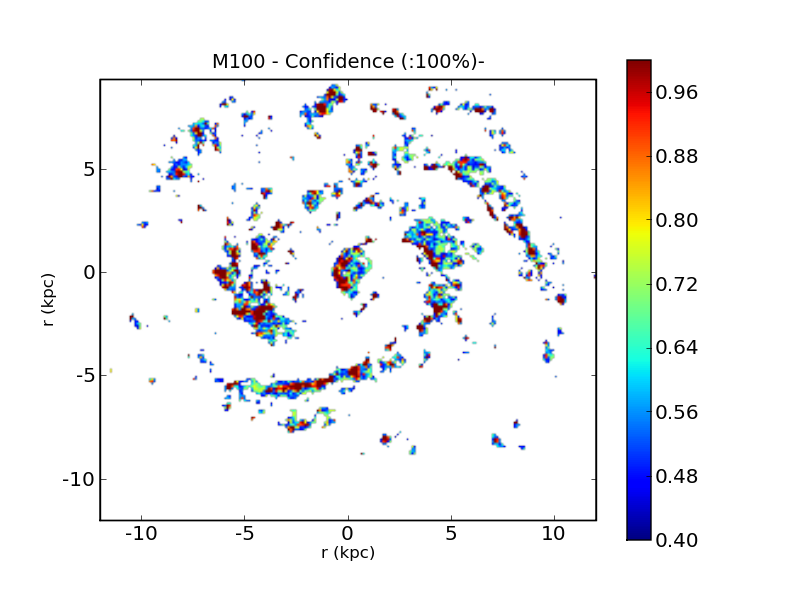}
\includegraphics[width=0.45\textwidth]{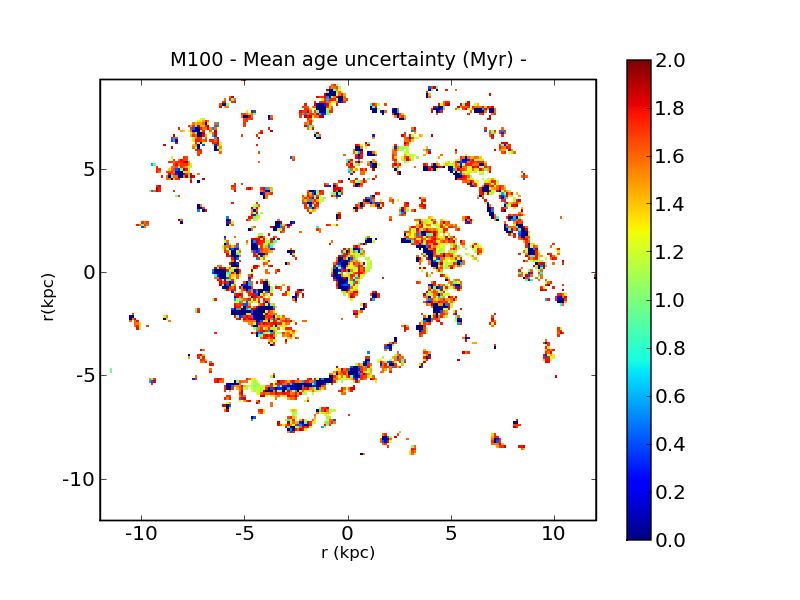}
\includegraphics[width=0.45\textwidth]{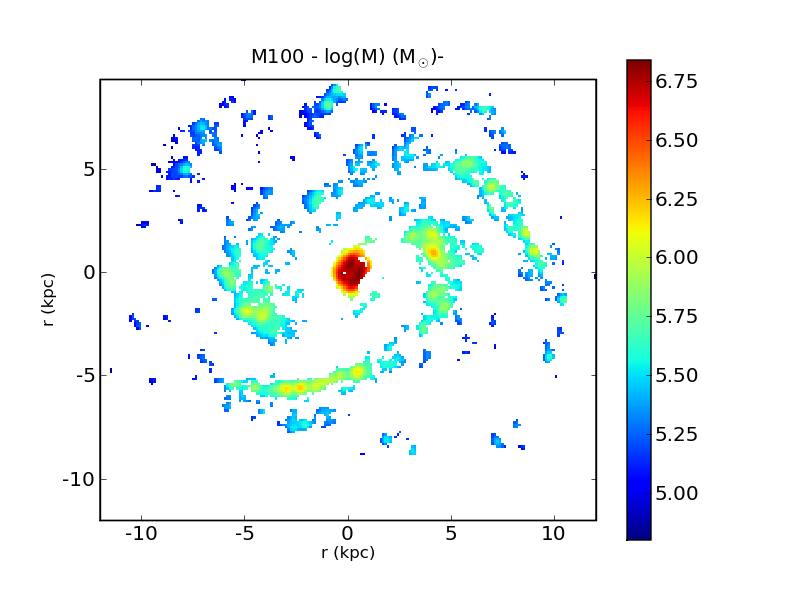}
\includegraphics[width=0.45\textwidth]{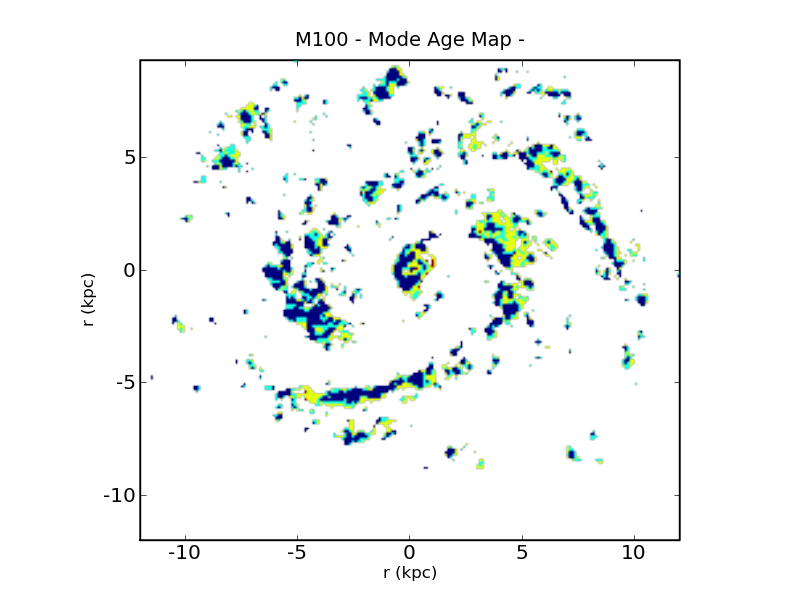}
\caption{Same as Fig. \ref{fig7} for the galaxy M100.}
 \label{fig9}
\end{figure*}

\begin{figure*}
\centering
\includegraphics[width=0.45\textwidth]{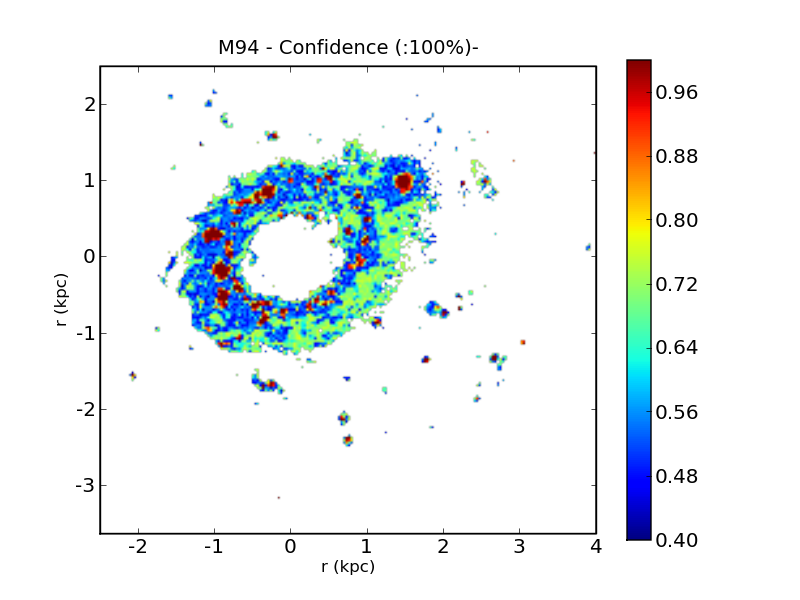}
\includegraphics[width=0.45\textwidth]{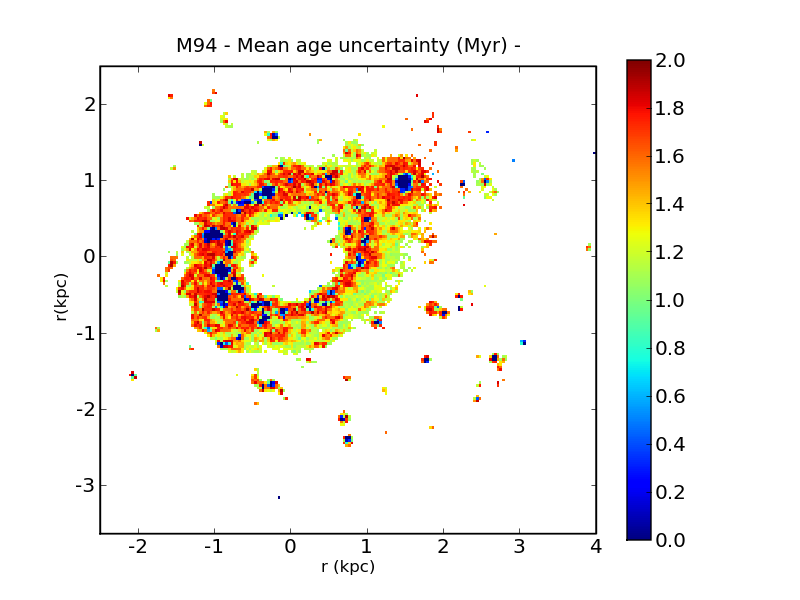}
\includegraphics[width=0.45\textwidth]{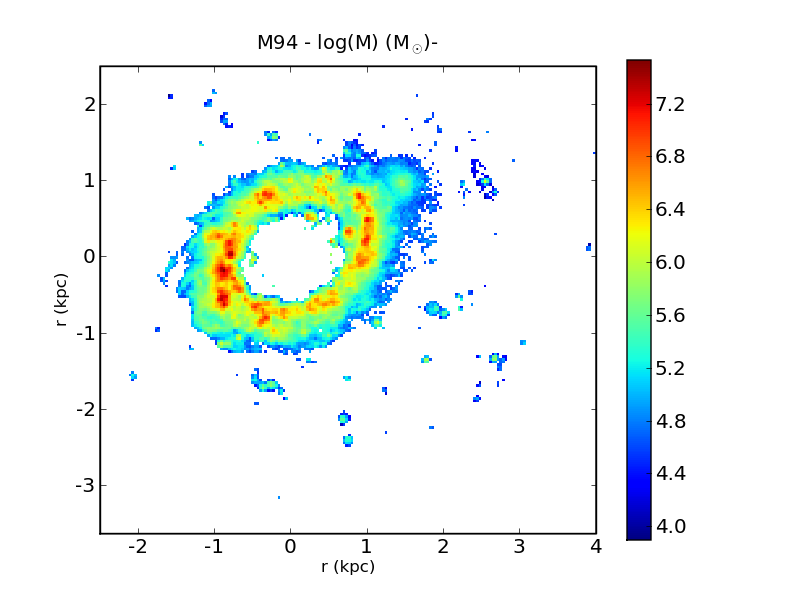}
\includegraphics[width=0.45\textwidth]{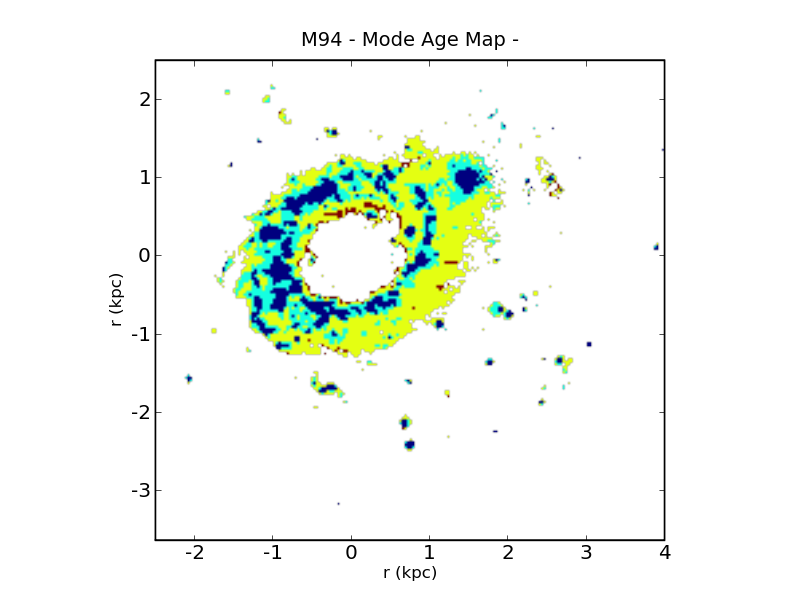}
\caption{Same as Fig. \ref{fig7} for the galaxy M94.}
 \label{fig10}
\end{figure*}

\begin{figure*}
\centering
\includegraphics[width=0.45\textwidth]{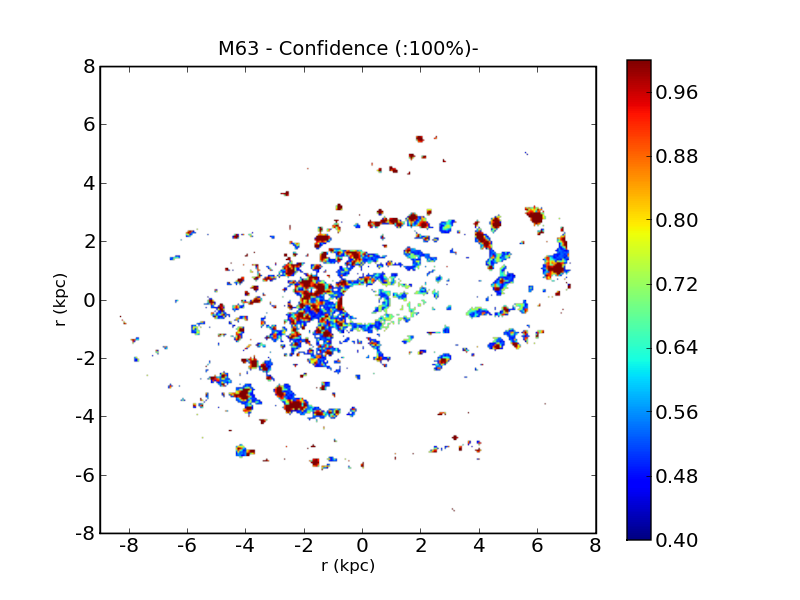}
\includegraphics[width=0.45\textwidth]{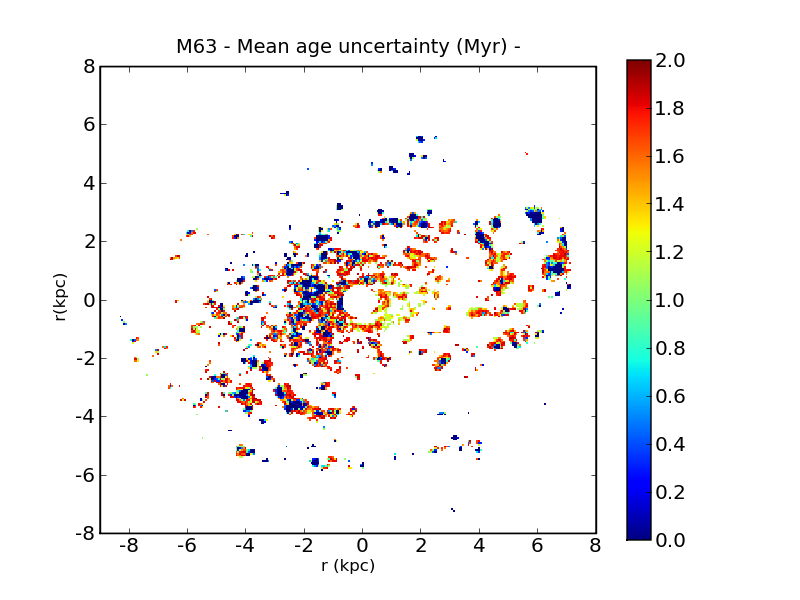}
\includegraphics[width=0.45\textwidth]{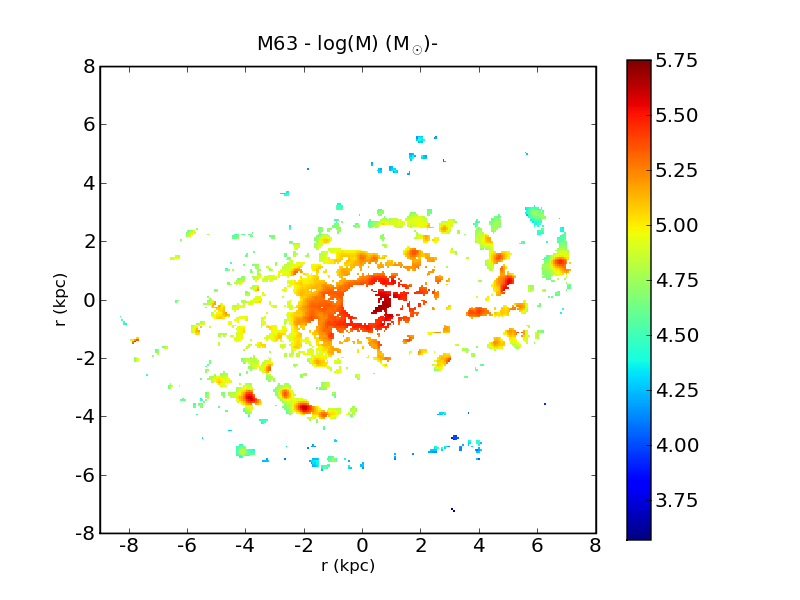}
\includegraphics[width=0.45\textwidth]{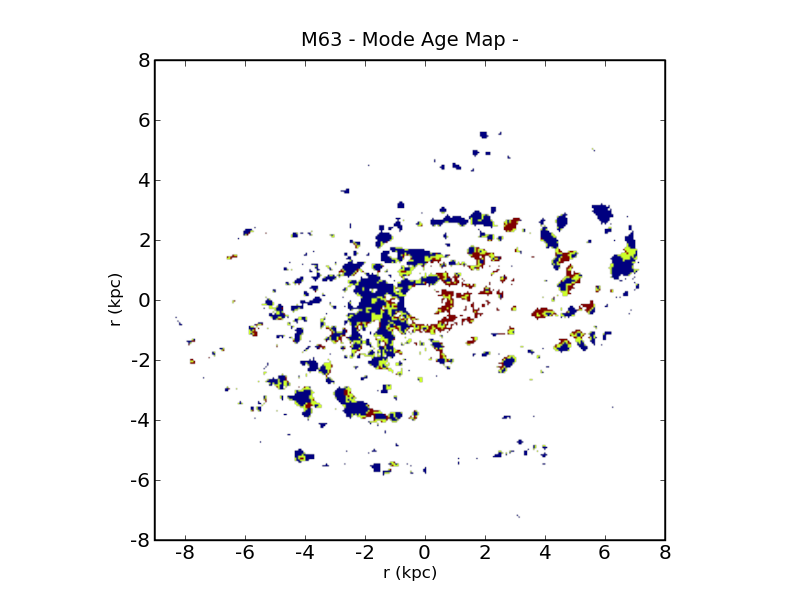}
\caption{Same as Fig. \ref{fig7} for the galaxy M63.}
 \label{fig11}
\end{figure*}

\begin{figure*}
\centering
\includegraphics[width=0.45\textwidth]{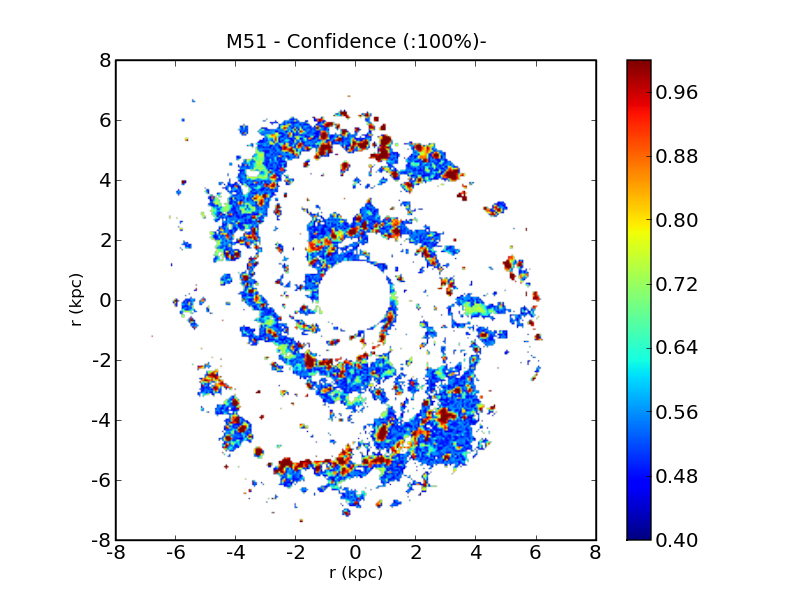}
\includegraphics[width=0.45\textwidth]{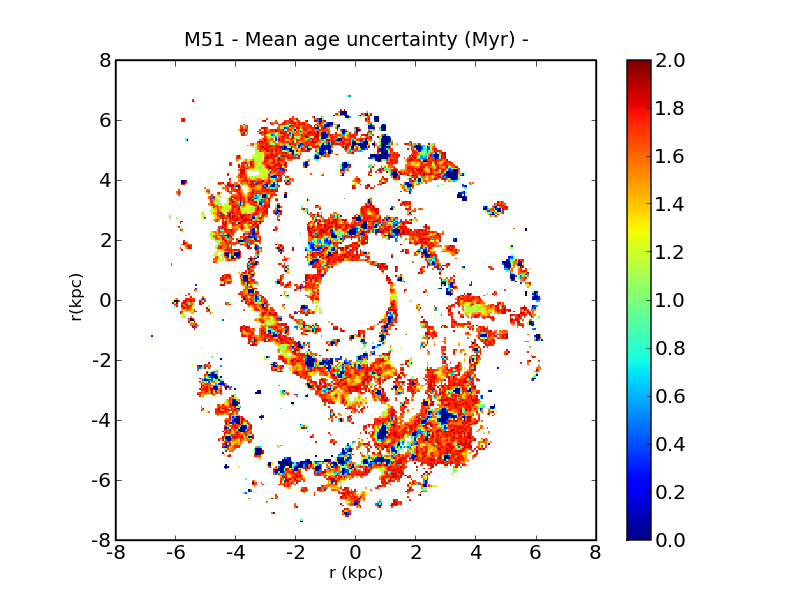}
\includegraphics[width=0.45\textwidth]{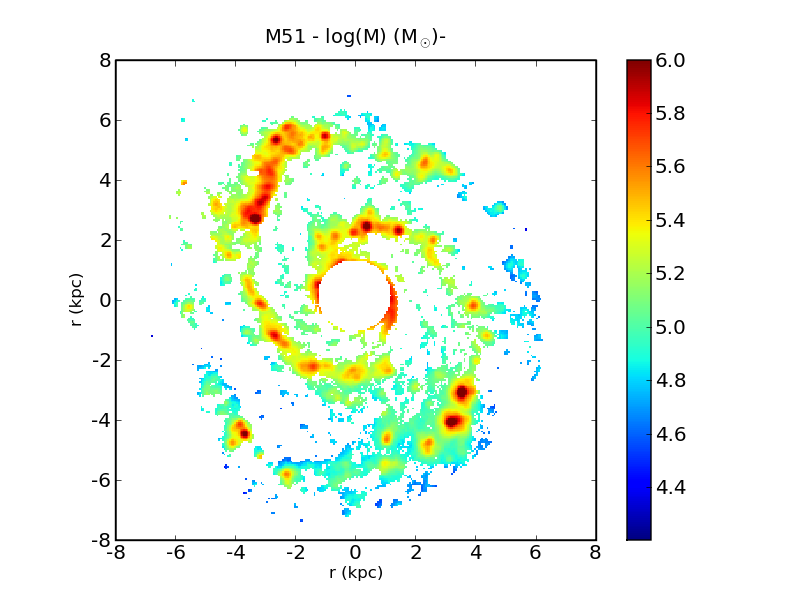}
\includegraphics[width=0.45\textwidth]{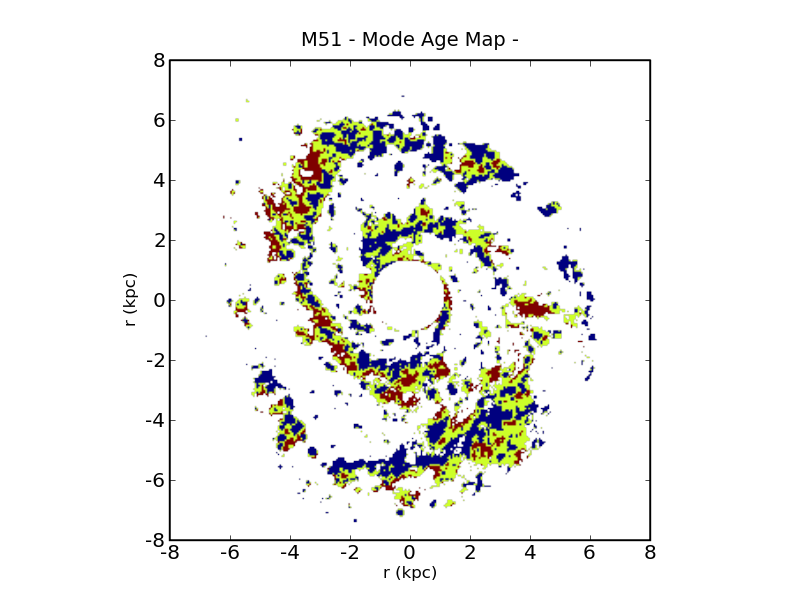}
\caption{Same as Fig. \ref{fig7} for the galaxy M51.}
 \label{fig12}
\end{figure*}


\section{Discussion}\label{discussion}


This work shows that a comparison of the \Ha\ and FUV observations of nearby spiral galaxies is a relatively direct way to probe burst age variations in spirals. Age gradients are common features along or across the arms of grand-design spiral galaxies, 
as well as in circumnuclear (M94) and in irregular \hii\ regions (IC 2574). In IC 2574, secondary star formation is observed on the periphery of the primary site of star formation, and probably triggered by the latter. For M63, the age gradient
occurs across the entire galactic disk, with the eastern side marked by a younger population ($<4$ Myr), while
in the inner $\pm3$ kpc of the western side an older population (6-9 Myr) dominates. Similarly in M74, there are age gradients across the spiral arms and an increasing age gradient from the inner to the outer parts of the galaxy. 
This corresponds with a very strong \Ha\ emission in the inner part of this galaxy. 

The age maps presented here provide a global view of the star formation processes taking place in the galactic disks.
They relate the maximum scale of coherent star formation to other large scale processes of star formation in galaxies, such as density waves and, in the case of M51, dynamics. However, it is beyond the scope of the present work to statistically 
favour one star formation mechanism over another given the extremely small sample size in hand.

One intriguing observational test recently proposed by Dobbs \& Pringle (2010) predicts a different distribution of young star cluster
ages depending on the mechanism for the excitation and maintenance of the spiral arms (see their Fig. 2).
They suggest methods for age-dating clusters in nearby galaxies as a means of distinguishing between the various theoretical 
models for spiral arm formation. They consider four canonical galaxy models: (i) fixed pattern speed, (ii) galaxy with a bar, 
(iii) a flocculent spiral, and (iv) a tidally induced spiral. For each model in turn, they estimate the expected distribution of 
star clusters for ages ranging from $\sim 2$ to 130 Myr.
Although our data only cover the most recent  star formation episodes up to 10 Myr, we find a spatial distribution of regions 
consistent with the predictions of Dobbs \& Pringle (2010), as well as a similar age distribution.
M74, and to a lesser degree M100, show the expected distribution for a spiral with a fixed pattern speed and/or a bar, with 
a monotonic sequence of ages across the spiral arms from young to old, as predicted by standard density wave theory.


Our age maps for M74, M100 and M51 show the expected distribution for a spiral with a
ﬁxed pattern speed and/or a bar, with a monotonic sequence of ages across the spiral
arms from youngest to oldest. Dobbs \& Pringle (2010) model M51 as a tidally induced
spiral undergoing a double interaction, and find that the age distribution of its
star clusters ($\sim$2 Myr to 130 Myr) do not show the same trend as we have found in
the younger age maps. Instead, their models show a rather a complex distribution,
where clusters of different ages appear simultaneously in the same region.
We infer that the youngest stellar populations show age gradients across the spiral arms, as predicted by standard spiral density wave theory.

In the cases of IC2574 and M94 it is difficult to identify a single possible mechanism for the triggering of star formation
in these galaxies. The age map for IC 2574 only shows local regions, like the giant northeastern \hii\ complex. Likewise 
for M94, the inner ring of intense active star formation dominates its age map. With observations of larger data sets the
method developed here will be a powerful tool when applied to larger, statistically-significant samples of nearby 
star-forming galaxies.


\section{Summary and Conclusions}\label{summary}


We have analyzed a sample of six spiral galaxies  to study the recent history
of their star formation.  The study of the spatial distribution of recent and ongoing star formation provides key
information about the evolutionary properties of gas-rich galaxies and the physical processes driving their evolution.

In this paper we have derived 2D age maps for each galaxy from a careful comparison of the UV and \Ha\ emission on a 
pixel-based basis, using Starburst 99 models to calibrate the relative fluxes in each band as function of age. While
the absolute ages derived from this method may carry significant uncertainties, the {\sl relative} ages derived through
this technique are relatively reliable and robust. 
The use of a pixel-wise age dating technique allows age mapping of the youngest stellar 
population  without prior assumptions about the spatial distribution of the star forming regions. The technique allows
the spatial characterization of the age distribution which for HII regions within a range of distance in the Local Volume 
through their spatially-integrated light. The far infrared flux is used to correct for extinction due to dust on pixel-scales
and the flux uncertainties in $F_{\rm H\alpha}$ and $F_{\rm FUV}$ are folded in throughout.

Derivations of star formation ages in this way are subject to a number of systematic factors that we have carefully
characterised to give confidence in our measurements. We have checked the validity of using a single stellar
population model in pixel-sized regions by ensuring that our pixel-sized masses are well in excess of the minimum
mass thresholds advocated by Cerviño et al. (2003). We have explored a range of different spatial binning scales
to verify that undersampling does not affect the spatial distribution of ages. Finally, we have also run our models 
over a range of plausible IMFs and metallicities to yield confidence bounds on our derived ages. As a result of these
tests, we have found that the average confidence in the age assignment is 64-70\%, although in excess of 80\%
of regions have confidence values equal to or greater than this. We also find that the average combined uncertainty 
on our ages is $1.4\pm0.5$ Myr, much less than the youngest age bin (4 Myr) adopted for our analysis.

The age maps for our six galaxies exhibit a range of characteristics. The grand design spirals M74 and M100
show evidence of age gradients along the spiral arms, with youngest to oldest running from centre to edge. This is 
consistent with star formation induced by spiral wave density theory. 
With its nearby companion, NGC 5195, M51 is typically modelled as a tidally induced spiral, undergoing a double interaction.
However, its age distribution of star clusters, from $\sim 2$ Myr to around 130 Myr, does not show a clear trend with respect to
its interacting companion and the most recent episodes of star formation. This lends support to the model
of Salo \& Laurikainen (1999) that sees NGC 5195 having already undergone multiple encounters with the
disk of M51, with no preferred specific site for recent star formation. Multiple encounters are also needed to explain
the long tidal tail of M51 to its companion and its characteristic kink.
The flocculent spiral M63 shows a gradient from edge to edge rather than radially,
suggesting that spiral density waves are not the dominant driver of star formation in such galaxies where the arms
are more loosely wound and less well-defined.

\section*{Acknowledgments}

MCSG thanks the Anglo-Australian Observatory for hospitality during stays in 2007 and 2008 during which time the
early stages of this work was completed. We also thank valuable input to this project from S. Cianci and M. Killedar
in its initial stages. This work is part of the PhD dissertation of M. Carmen Sánchez-Gil, funded by the Spanish Ministerio 
de Ciencia e Innovaci\'on (MICINN), under the FPU grant AP-2004-2196. We acknowledge financial support from Spanish MICINN through grants AYA2007-64052 and AYA2007-64712 and from Consejer\'{\i}a de Educaci\'on y Ciencia (Junta de Andaluc\'{\i}a) through TIC-101, TIC-4075 and TIC-114.

\newpage

\appendix

\section{IRX-$\beta$ relation}\label{IRX-beta}

The IRX-$\beta$ relation (Heckman et al. 1995, Meurer et al. 1999) relates the ultraviolet slope, $\beta$ to
the ratio of total infrared to UV flux, (defined as $IRX$ = $F_{TIR}/F_{FUV}$). It demonstrates that the
UV light absorbed by the dust is re-emitted as thermal emission in the far infrared, thereby allowing one
to estimate the extinction in the UV from the slope. Here we independently derive our own IRX-$\beta$
relation based on our pixel-based analysis of the six star forming galaxies in our sample.

Values of the UV spectral slope were derived per pixel for each of the six galaxies according to
the definition of Kong et al. (2004),

\begin{equation}
\beta = \frac{log(f_{FUV})-log(f_{NUV})}{log(\lambda_{FUV})-log(\lambda_{NUV})}
\label{eqKong}
\end{equation}

where $f_{FUV}$ and $f_{NUV}$ are the far- and near-UV fluxes from the respective FUV and NUV GALEX filters
(in ergs$^{-1}$cm$^{-2}$\AA$^{-1}$).
The central wavelengths of each bandpass are $\lambda_{FUV}$ = 1520 \AA, $\lambda_{NUV}$ = 2310\AA.
The total far-infrared fluxes were calculated as described in Sect \ref{SData}.

Cortese et al. (2006) use a linear fit between the log(IRX) and $\beta$, 
\begin{equation}
log(IRX)= a+b\beta
\label{eqIRX1}
\end{equation}
although other authors (Kong et al. 2004, Boissier et al. 2007)
have found that adding a non-linear term 
\begin{equation}
IRX = 10^{a+b\beta} - c
\label{eqIRX2}
\end{equation}
produces a better fit to the data. We have fitted both linear and non-linear relations to our log(IRX)--$\beta$ relation
and compared our results to these authors. Outliers were rejected if they were beyond 4-standard deviations of the mean.
The linear fit was calculated using a classical ordinary least squares fit (OLS) of log(IRX) on $\beta$ and vice versa. 
The final fit was taken as the bisector of the two OLS lines and is the fit is similar to that given by Cortese et al. (2006)
for an optically-selected sample of normal star-forming galaxies in nearby clusters. The upper left panel of 
Fig. \ref{fig13} shows our fits to the full sample compared to those of Cortese et al. (2006) and the fit parameters are
summarised in Table~\ref{tab4}.

The upper right panel of Fig. \ref{fig13} shows our non-linear fit to the same data compared to those of Kong et al. (2004) and
Boissier et al. (2007). Our fit is closer to that of Kong et al. but shifted to higher values of $\beta$ and with smaller slope,
Kong et al. use a sample of 50 starburst galaxies, while Boissier et al. take a sample of 43 nearby, late-type galaxies. With
the exception of the interacting M51, it can be argued that all the galaxies in our sample are normal quiescent star forming 
systems. However,  because we consider emission on pixel scales (rather than on a galaxy-wide scale),  each pixel ought
to be considered as a \hi\i\ region, in which case we would expect the behaviour to be closer to the starburst sample
fitted by Kong et al., as is seen in Fig. \ref{fig13}. Table~\ref{tab4} gives the parameters of each non-linear fit.

We use the Akaike Information Criterion (AIC; Akaike 1974) to choose the better fit between the linear and non-linear cases,
defined as

\begin{equation}
AIC = -2 \ln(L) + 2K
\label{eqAIC}
\end{equation}
where $\ln(L)$ is the logarithm of maximized likelihood function, and $K$ is the number of parameters in the model. 
In the case of least squares regression for normally distributed errors, (as it is the case here), it can be computed as
\begin{equation}
AIC = N \log(\hat{\sigma}^2) + 2K
\label{eqAIC2}
\end{equation}
where $\hat{\sigma}^2$ is the variance and $N$ is the sample size. 
The `best' model is that with the lowest AIC among a set of specified models,
and which best describes the data with the fewest number of free parameters.
Applying equation (\ref{eqAIC2}) to the fits from equations (\ref{eqIRX1}) and (\ref{eqIRX2}) reveal the non-linear relation
to be the better fit of the two.

The pixel points in the upper left panel of Fig. \ref{fig13} are colour-coded by galaxy and highlight the different regions
occupied by each galaxy in the IRX--$\beta$ plane. The points of the irregular galaxy IC 2574 are confined to lower IRX 
and $\beta$ values in contrast with the other spiral galaxies. We note that M51 exhibits the greatest scatter about its
IRX--$\beta$ fit compared to the other spirals. 

The lower panels of Fig. \ref{fig13} show the curves of best fit for each galaxy compared to the overall fit for the sample, in
both the linear (left panel) and non-linear (right panel) cases. While the non-linear case is a better fit to the sample as
a whole, it would appear to be inappropriate in the case of individual galaxies. Notwithstanding our small sample size, 
the upper left panel of Fig. \ref{fig13} suggest that different morphologies are reflected in varying ranges for
IRX and $\beta$, and would explain the non-linear nature of the full sample. We would conclude that for a sufficiently 
restricted mix of morphological types and galaxy mass, the IRX--$\beta$ relation is linear, and that care should be
exercised when fitting a sample that is not.

Table \ref{tab5} lists the fit parameters to individual galaxies for both the linear and non-linear case.

\begin{table*}
\caption{IRX--$\beta$ fits for the full sample}
\begin{tabular}{ @{}l@{} l l l@{}}
\hline
Fit & & This paper & Literature \\
\hline
& OLS(Y$|$X) & log(IRX)  $=$  0.459($\pm$0.002)$\beta$ $+$ 1.001($\pm$0.002), $R=0.662$&  \\
Linear & OLS(X$|$Y) & $\beta$ $=$ 1.033($\pm$0.006)$log(IRX)$ $-$ 1.387($\pm$0.005), $R=0.662$ & log(IRX)  $=$  0.7($\pm$0.06)$\beta$ $+$ 1.30($\pm$0.06)$^{{\bf (a)}}$ \\
& OLS bisector & log(IRX)  $=$  0.662($\pm$0.004)$\beta$ $+$ 1.151($\pm$0.004) & \\
\hline
Non-linear &\multicolumn{2}{c}{ IRX $=$ 10$^{1.225(\pm0.009)+0.239(\pm0.004) \beta}$ $-$ 3.736($\pm$0.311), $R=0.589$} & IRX $=$10$^{2.10+0.85 \beta}$ $-$ 0.95$^{{\bf (b)}}$ \\
&&& IRX $=$10$^{1.145+0.324 \beta}$ $-$ 3.136$^{{\bf (c)}}$\\
\hline
&&&\\
\end{tabular}
\begin{flushleft} 
${}^{{\bf (a)}}$Cortese et al. 2006 \\
${}^{{\bf (b)}}$Kong et al. 2004 \\
${}^{{\bf (c)}}$Adapted from Boissier et al. 2007; The fit given is IRX $=$10$^{0.561+0.713 (FUV-NUV)}$ $-$ 3.136, where FUV$-$NUV is the color between the two GALEX bands. The direct relation between $\beta$ and the (FUV-NUV)  is given by equation \ref{eqKong}.
\end{flushleft}
\label{tab4}
\end{table*}%

\begin{table*}
\caption{IRX--$\beta$ fits by galaxy}
\begin{tabular}{ c c c c | c c c c}
\hline
&\multicolumn{3}{c|}{Linear fit: log(IRX)$= \tilde{a}\beta+\tilde{b}$} &\multicolumn{4}{c}{Non-linear fit: IRX$=$10$^{a+b \beta} -$ c} \\
&$\tilde{a}$&$\tilde{b}$&$$R~~~~&$a$&$b$&$c$&$R$ \\
\hline
M51&0.575$\pm$0.007&1.155$\pm$0.010&0.493~~~~& 1.292$\pm$0.031&0.169$\pm$0.013& 6.245$\pm$1.36&0.412\\
M63&0.409$\pm$0.008&1.191$\pm$0.009&0.659~~~~& 1.107$\pm$0.027&0.288$\pm$0.014& -3.687$\pm$0.72&0.639\\
M74& 0.623$\pm$0.015 &1.090$\pm$0.014 & 0.357~~~~& 0.631$\pm$0.078 &0.223$\pm$0.054 &-1.739$\pm$0.75 & 0.276\\
M94&0.564$\pm$0.009&1.156$\pm$0.011&0.673~~~~& 1.241$\pm$0.033&0.185$\pm$0.016&5.79$\pm$1.28&0.672\\
M100&0.590$\pm$0.011&1.070$\pm$0.001&0.674~~~~& 1.099$\pm$0.023&0.305$\pm$0.014&1.518$\pm$0.59&0.693\\
IC 2574&0.835$\pm$0.016&1.042$\pm$0.043&0.44~~~~& 0.231$\pm$0.026&0.245$\pm$0.054&0.112$\pm$0.16&0.359\\
\hline
\end{tabular}
\begin{flushleft} 
\end{flushleft}
\label{tab5}
\end{table*}%

\begin{figure*}
\includegraphics[width=0.45\textwidth]{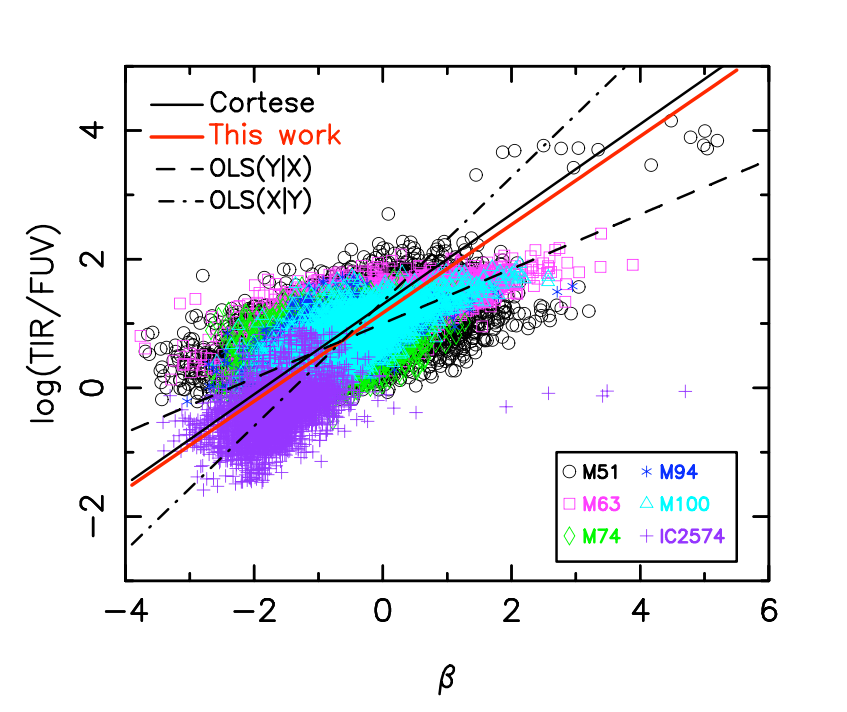}
\includegraphics[width=0.45\textwidth]{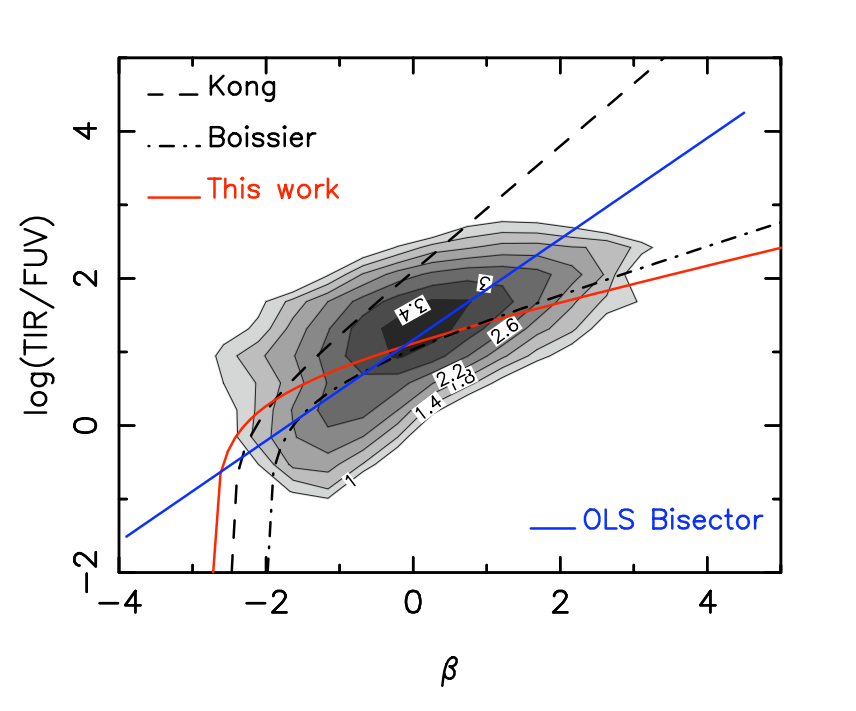}
\includegraphics[width=0.45\textwidth]{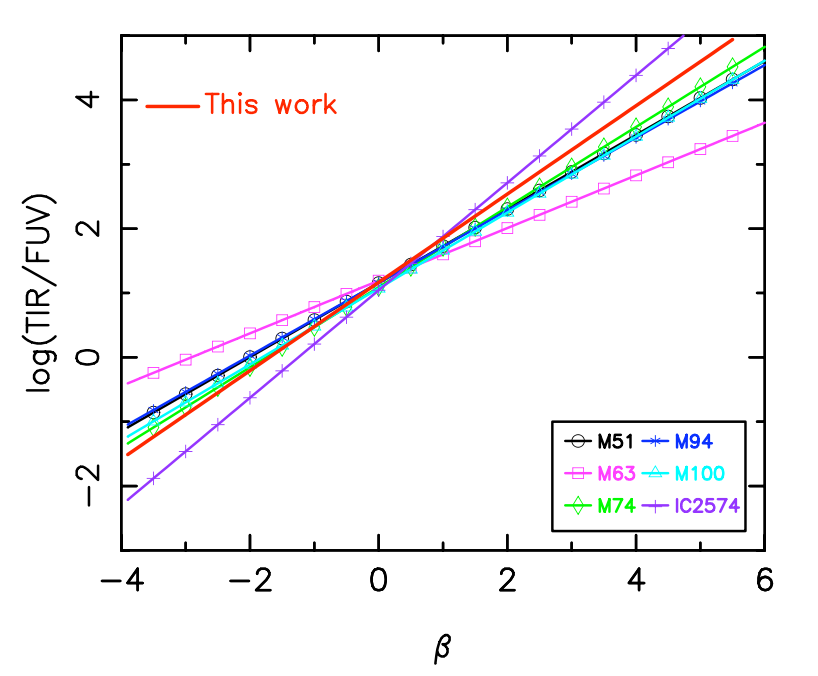}
\includegraphics[width=0.45\textwidth]{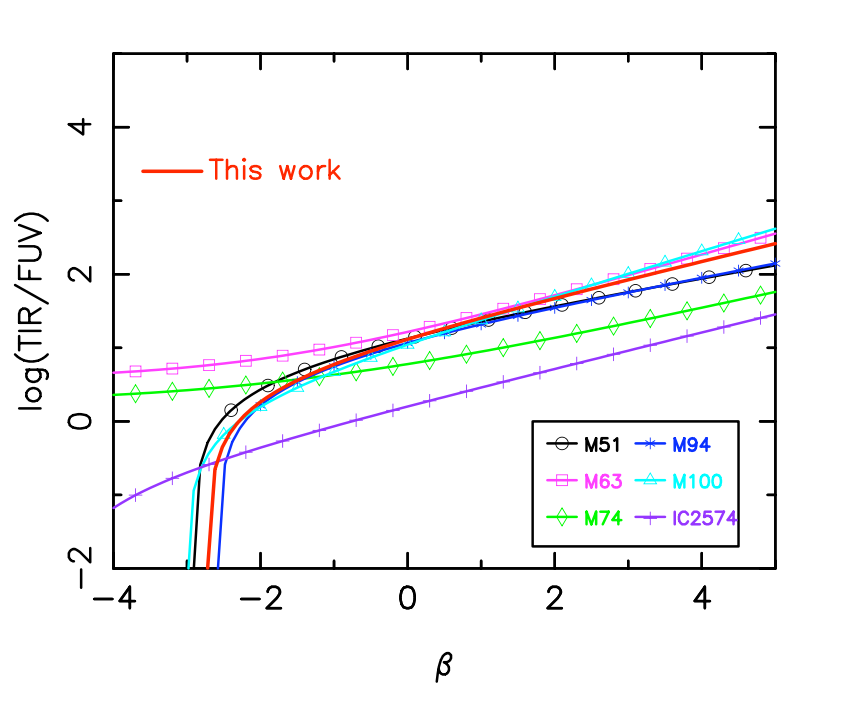}
\caption{IRX--$\beta$ plots where each galaxy is represented by a different colour and symbol, as specified in the key.
{\em upper left:}  Linear fit for all points combined (red line) compared to that of Cortese et al. (2006) (black line). Also
shown are the OLS (Ordinary Least Squares) bisectors OLS(Y$|$X) and OLS(X$|$Y) from which our full fit was derived.
{\em upper right:} Non-linear fit for all points combined (red line) compared to those of Kong et al. (2004) (dashed line) and 
Boissier et al. (2007) (dot-dash line). In this case we have plotted our full sample as a density distribution with log contours.
Also shown (blue line) is the linear fit from the upper left panel.
{\em lower left:} Linear fits of the IRX--$\beta$ relation for individual galaxies, using identical colours and symbols to those in
the upper left panel.
{\em lower right:} Non-linear fits per galaxy showing the same colours and symbols as the other panels. } 
\label{fig13}
\end{figure*}

\bsp

\label{lastpage}

\end{document}